\documentclass[10pt,preprint2]{aastex}
\newcommand{\alpht  }{\alpha_\mathrm{t}}
\newcommand{\alphs  }{\alpha_\mathrm{s}}
\newcommand{\alphOV }{\alpha_\mathrm{OV}}
\newcommand{\alphOM }{\alpha_\mathrm{OM}}
\newcommand{\alphMLT}{\alpha_\mathrm{MLT}}
\newcommand{\Mstar  }{\,{M}_{*}}

\newcommand{\Msun   }{{M}_{\sun}}
\newcommand{\Rsun   }{{R}_{\sun}}
\newcommand{\Lsun   }{{L}_{\sun}}
\newcommand{\km     }{\mbox{km}}
\newcommand{\Mfit   }{M_\mathrm{fit}}
\newcommand{\Teff   }{T_\mathrm{eff}}
\newcommand{\lz     }{l=0}
\newcommand{\lo     }{l=1}
\newcommand{\ltw    }{l=2}
\newcommand{\ltr    }{l=3}
\newcommand{\Delnunl}{\Delta \nu(n,l)}
\newcommand{\delnunl}{\delta \nu(n,l)}
\newcommand{\BruntV }{{Brunt-V{\"a}is{\"a}l{\"a} }}
\newcommand{\ProcA  }{{Procyon~A }}
\newcommand{\ProcAk }{{Procyon~A}}
\newcommand{\ProcAs  }{{Procyon~A's }}
\newcommand{\Fig    }{Figure~}
\newcommand{\Figs   }{Figures~}
\newcommand{\Sec    }{Section~}

\newcommand{\Eqn    }{equation~}
\newcommand{\eq     }{eq.~}

\newcommand{\Hp     }{H_{\!{P}}}
\newcommand{\Cp     }{C_{\!{P}}}
\newcommand{\Pt     }{P_\mathrm{t}}
\newcommand{\jt     }{j_\mathrm{t}}
\newcommand{\jw     }{j_{\wturb}}
\newcommand{\Frad   }{F_\mathrm{rad}}
\newcommand{\krad   }{k_\mathrm{rad}}
\newcommand{\delM   }{\Delta m}
\newcommand{\betr   }{\beta_{r}}

\newcommand{\nablmu }{\nabla_{\!{\mu}}}
\newcommand{\nablad }{\nabla_{\!\mathrm{ad}}}
\newcommand{\nablrad}{\nabla_{\!\mathrm{rad}}}
\newcommand{\Pgas   }{P_\mathrm{gas}}
\newcommand{\rcz    }{r_\mathrm{cz}}

\newcommand{\rb     }{r_\mathrm{b}}
\newcommand{\wturb  }{\bar{\omega}}
\newcommand{\punkt  }{\,\,\mbox{.}}
\newcommand{\lderivt}[1]{\frac{\mbox{D} #1}{\mbox{D}t}}
\newcommand{\gradm  }[1]{\frac{\partial #1}{\partial m}}

\newcommand{\gradn  }[1]{\frac{\partial #1}{\partial n}}

\slugcomment{Submitted to Astrophysical Journal -- Revision 1.}

\shorttitle{Core Overshoot: Improved Treatment and Constraints}
\shortauthors{Straka, Demarque, \& Guenther}

\begin{document}

\title{Core Overshoot: An Improved Treatment and Constraints from Seismic Data}

\author{Christian W. Straka\altaffilmark{1}}
\affil{Institut f\"ur Theoretische Astrophysik, Universit\"at Heidelberg, Tiergartenstra{\ss}e 15,\\
       69121 Heidelberg, Germany}
\email{cstraka@ita.uni-heidelberg.de}

\author{Pierre Demarque}
\affil{Department of Astronomy, Yale University, P.O. Box 208101, New Haven, CT 06520-8101}
\email{demarque@astro.yale.edu}

\and

\author{D. B. Guenther}
\affil{Department of Astronomy and Physics, Saint Mary's University, Halifax, N.S., Canada, B3H 3C3}
\email{guenther@ap.stmarys.ca}

\altaffiltext{1}{Visiting Astronomer, Department of Astronomy, Yale University}

\begin{abstract}
We present a comprehensive set of stellar evolution models
for \ProcA in an effort to guide future measurements of both
traditional stellar parameters and seismic frequencies
towards constraining the amount of core overshoot in \ProcA
and possibly other stars.  Current observational
measurements of \ProcA when combined with traditional
stellar modeling only place a large upper limit on overshoot
of $\alphOV < 1.1$. By carrying out a detailed pulsation
analysis, we further demonstrate, how p- and g-mode averaged
spacings can be used to gain better estimates of the core
size. For both p- and g-modes, the frequency spacings for
models without overshoot are clearly separated from the
models with overshoot. In addition, measurements of the
$\lz$ averaged small p-mode spacings could be used to
establish \ProcAs evolutionary stage.  For a fixed
implementation of overshoot and under favorable
circumstances, the g-mode spacings can be used to determine
the overshoot extent to an accuracy of $\pm
0.05\,\Hp$. However, we stress that considerable confusion
is added due to the unknown treatment of the overshoot
region. This ambiguity might be removed by analyzing many
different stars.  A simple non-local convection theory
developed by Kuhfu{\ss} is implemented in our stellar
evolution code and contrasted with the traditional
approaches. We show that this theory supports a moderate
increase of the amount of convective overshoot with stellar
mass of $\Delta\alphOV \simeq + 0.10$ between $1.5\,\Msun$
and $15\,\Msun$.  This theory places an upper limit on
\ProcAs core overshoot extent of $\sim 0.4\,\Hp$ which
matches the limit imposed by Roxburgh's integral criterion.
\end{abstract}


\section{Introduction}

One of the most important deficiencies in stellar evolution theory
is the lack of a physically correct treatment of convective motions and
convective transport of energy. The extreme difficulty of
solving the well-known hydrodynamic equations on all different scales
is a long standing mathematical and computational challenge.
Due to these unsolved complications of an accurate treatment,
convection in stars has been described by mixing length theory
(MLT) \markcite{1958ZA.....46..108B}({B{\" o}hm-Vitense} 1958),
a simple phenomenological method for treating convection locally.
Within the framework of MLT
\markcite{1965MNRAS.130..223R}{Roxburgh} (1965) and \markcite{1965ApJ...142.1468S}{Saslaw} \& {Schwarzschild} (1965)
were the first who estimated the
penetration of convective motions into the radiative regions due
to a finite convective velocity at the Schwarzschild boundary defined
by vanishing acceleration through buoyancy. The result, a negligible
amount of overshoot of $10^{-2}\Hp$ was criticized by
\markcite{1973ApJ...184..191S}{Shaviv} \& {Salpeter} (1973) who found appreciable overshoot
within the framework of a simplified non-local mixing length model.
Most stellar modelers since have considered overshoot only in a
parameterized way by extending the core size by some fraction
$\alphOV$ of the pressure scale height $\Hp$ taken at the locus
of the Schwarzschild boundary.

In the present work we implement a physically better
motivated phenomenological model for convection including
overshoot into stellar evolution calculations,
a {\it non-local} convection theory developed
by \markcite{1986A&A...160..116K}{Kuhfu{\ss}} (1986) and we contrast
this approach with the traditional ones. In both cases we are
left with free parameters. In MLT only the overshoot parameter
$\alphOV$ has to be considered for the extension
of the convective cores. The situation is more
complicated for the \markcite{1986A&A...160..116K}{Kuhfu{\ss}} (1986)
convection theory (KCT).
Nonetheless, we vary only one parameter $\alpht$ that is 
the most relevant for the amount of overshoot.

Empirically, isochrone fitting to color-magnitude diagrams (CMDs)
of open clusters have so far given the most quantitative
results for the amount of core overshoot. The underlying
hypothesis in these studies is that all stars have been formed
from the same material at the same epoch. Both age and metallicity
of the cluster are not known and a range of parameters has to be
matched to observations. However, most uncertainty is introduced
through contamination of unresolved binary systems and
variable stars, the need to clean the sample from non-member
field stars \markcite{1997AJ....113.1045K}(cf., {Kozhurina-Platais} {et~al.} 1997),
the relatively small number of stars in the sample and
the systematics introduced by transforming the CMD to the theoretical HRD
diagram. Studies of this kind have been performed to
test the overshoot parameter, e.g., by \markcite{1974ApJ...193..109P}{Prather} \& {Demarque} (1974),
\markcite{1981A&A....93..136M}{Maeder} \& {Mermilliod} (1981), \markcite{1998MNRAS.298..525P}{Pols} {et~al.} (1998) and
\markcite{1994ApJ...426..165D}{Demarque}, {Sarajedini}, \&  {Guo} (1994) who obtain for their best fit a
value of $\alphOV = 0.23$, which can be regarded as a canonical value.

In a different approach, detached and evolved
eclipsing binaries have been used to constrain
the amount of overshoot. \markcite{2000MNRAS.318L..55R}{Ribas}, {Jordi}, \& {Gim{\'  e}nez} (2000) studied
eight such objects and found an overall best fit of
$\alphOV = 0.25 \pm 0.05$ and more strikingly a clear trend
of increased overshoot with increasing mass.

%Aerts reference here
Recently, \markcite{2003Sci...300.1926A,2004A&A...415..251D}{Aerts} {et~al.} (2003); {Dupret} {et~al.} (2004)
have used observed seismic constraints to deduct the extent
of core overshoot in the $\beta\,$Cep star HD 129929. Based
on fitting three identified pulsation frequencies
these authors claim an overshoot parameter of $\alphOV=0.10\pm0.05$
for this $\sim 9\,\Msun$ star.
While unambiguously identified single frequencies
hold the promise of strongly constraining stellar models, we think
that many more frequencies than free stellar parameters must be used
for such an analysis. We show in this paper, that the frequencies
are also influenced by the adopted overshoot prescription, which
constitutes an additional uncertainty not previously taken into
account in stellar modeling.

%new direction
Our main objective in this study is to show how averaged
frequency spacings can be used to constrain core overshoot
in carefully selected single stars like \ProcA. The
advantage of this technique is that averaged spacings may
not depend on individual mode identifications as critically
as single modes do. Results from this approach produce
weaker constraints on stellar models but on the other hand
should yield more solid results.  By extensively analyzing
the pulsation properties of many different models for \ProcA
we show that the oscillation characteristics contain very
detailed information about overshoot and the stellar
interior. We outline ways to constrain the overshoot
parameter that may become practical in the future.

\section{Stellar Evolution Theory}

\subsection{General Input Physics}
We use the Yale Rotating Stellar Evolution Code (YREC)
\markcite{1988PhDT........14P,1992ApJ...387..372G}({Pinsonneault} 1988; {Guenther} {et~al.} 1992)
in its non-rotating configuration. The four classic stellar structure
equations, namely Poisson's, hydrostatic, energy transport and energy
equations are solved with a Henyey relaxation scheme in the interior and
a shooting technique in the outer envelope and atmosphere. The fitting
point of the interior model to the envelope is set at the fixed mass
$\Mfit = 10^{-12}\Mstar$. A nuclear reaction network is
solved separately. The separation of the chemical part from the structural
part can be done safely in the main burning stage of hydrogen burning
considered here since the nuclear timescale and therefore the change in
composition is long compared to the thermal and dynamical timescales
involved. \ProcA has a very thin outer convective zone of $\sim 10^{-5}\Mstar$
and in these outer layers standard MLT
is used.
%with a mixing length parameter of
%$\alphMLT=1.6$ which was calibrated in a standard solar model.

In the absence of overshoot, core convection is also treated in a standard
way by determining the regions of convective instability with the
Schwarzschild criterion. In regions where it
indicates convection, the temperature gradient in the energy transport
equation is set to the adiabatic gradient, thereby assuming fully developed
turbulent convection. The superadiabaticity which can be estimated to
be typically on the order of $10^{-6}$ in the deep stellar layers
is thereby neglected and the mixing length parameter present in MLT
becomes irrelevant.  The deviation of $10^{-6}$ from strict
adiabaticity is too tiny to have any effect on the stellar structure or
the pulsation frequencies which are mainly sensitive to the sound speed and
the mean molecular weight gradient. In addition, chemical species are mixed
instantaneously within convective regions justified by the short mixing
timescale compared to the nuclear timescale under the conditions
prevalent in this study.

For the material functions, the
OPAL equation of state \markcite{1996ApJ...456..902R}({Rogers}, {Swenson}, \&  {Iglesias} 1996) and
OPAL opacities \markcite{1996ApJ...464..943I}({Iglesias} \& {Rogers} 1996) for temperatures
above $\log T > 4.12$ are used, and in the low
temperature regime $\log T < 4.0$, we use opacities by
\markcite{1994ApJ...437..879A}{Alexander} \& {Ferguson} (1994). In the transition region the
low temperature opacity tables and the interior opacities are ramp
averaged.

The number of grid points can be crucial for calculating correct
convective core sizes and pulsation frequencies. YREC inserts
and deletes grid points according to specified criteria on structural variables
(e.g., pressure). With the help of some test calculations we find that
$\sim 900$ grid points are sufficient for both core size and
pulsation frequencies. Our models generally consist of about 1800 grid
points, twice as many as minimally required. 
The innermost mass shell is taken at $m_{1} = 10^{-5}\Mstar$
and we verified that this is about an order of magnitude
lower than needed to produce correct pulsation frequencies.

A note on terminology. \markcite{1991A&A...252..179Z}{Zahn} (1991)
proposes to distinguish between the term \emph{overshooting} and
\emph{penetration} since they have a more precise meaning in fluid
dynamics. In this nomenclature, \emph{overshooting} is reserved for
inefficient penetration, that does not alter the stable
temperature gradient whereas \emph{subadiabatic penetration} refers to
convective heat transport efficient enough to establish a nearly
adiabatic temperature gradient. In this paper, we prefer to
use \emph{overshoot} as the superordinate term
for all penetration beyond the stability boundary regardless of
its efficiency. However, we refer to penetration that does not
alter the temperature gradient as \emph{overmixing}.

\subsection{Overshoot Prescriptions}
\label{sec:ov}

There exists vast theoretical and experimental evidence for
overshoot in geophysical fluids, laboratory experiments
and computer simulations \markcite{1991A&A...252..179Z}({Zahn} 1991). With a simple
theoretical model utilizing the usual scaling laws of thermal convection
\markcite{1991A&A...252..179Z}{Zahn} (1991) demonstrates
that the subadiabatic penetration above a convective core should
amount to a substantial fraction of the core radius in the interior
of stars. If correct, overshoot would strongly influence
the evolution of stars for
its structure, chemical enrichment, lifetime and possibly its fate.

It has long been recognized that MLT, still commonly
used in almost all stellar evolution calculations, being a strictly
local theory, cannot account for the phenomenon of penetrative
motions beyond the classical boundary of convective cores. Many non-local
convection theories available from geophysics or other fields have been
largely ignored for two reasons. First, many of them are difficult to
implement within the numerical scheme adopted in stellar evolution and
probably more importantly contain short lived timescales of the order
of the \BruntV frequency, thus rendering it impossible to follow
stellar evolution on timescales comparable to the thermal and nuclear
timescales. Secondly, many of those theories contain more than one free
parameter giving argument to the view that more elaborate descriptions
may be more correct but nonetheless exhibit arbitrary degrees of freedom.
In this spirit, and
partly supported by the success of the one parameter MLT, all actual
stellar evolution calculations employing overshoot, extend
the boundary of the convective core by some fraction $\alphOV$ of the
local pressure scale height taken at this very boundary. The extended
region is treated differently by different authors, some only taking
additional mixing of chemical species into account while others
also alter the temperature gradient in this region. While this
approach being simple, it cannot be regarded as an adequate method to
capture the physical phenomenon underlying core overshoot and it is
not surprising that the value of $\alphOV$ is disputed not only over
its correct value, but also whether the same value should be taken
at the bottom and the top of convective regions, and whether higher
values should be used for stars with higher mass
\markcite{2000MNRAS.318L..55R}({Ribas} {et~al.} 2000).

We feel it is time to advance the prescription of overshoot
employed in stellar evolution calculations. We choose to implement
a one dimensional non-local convection
theory developed by \markcite{1986A&A...160..116K}{Kuhfu{\ss}} (1986) in the simplest
possible way. Note that \markcite{1986A&A...160..116K}{Kuhfu{\ss}} (1986) derives
an equation for the specific turbulent kinetic energy by proper spherically
averaging of the first order perturbed Navier-Stokes equations. The
principal difficulty that arises in all phenomenological theories
of convection is to model the (unknown) correlation functions of
the fluctuating quantities. This is done within the framework of
anelastic and diffusion type approximations by introducing free parameters
for every term arising in the equation. Thus, we are left with five
free parameters. \markcite{1986A&A...160..116K}{Kuhfu{\ss}} (1986) fixes two of them
by matching the convective velocity and the convective flux
with the corresponding MLT values. We do not alter those
in this study. The remaining parameters consist of a
mixing length parameter $\alphMLT$ (analogous to MLT) and an
overshoot parameter $\alpht$ (analogous to $\alphOV$).
The third free parameter does not
appear in the original \markcite{1986A&A...160..116K}{Kuhfu{\ss}} (1986)-theory and
was first introduced in it by \markcite{1998A&A...340..419W}{Wuchterl} \& {Feuchtinger} (1998), a parameter
$\betr$ that essentially limits the local pressure scale height $\Hp$
to a more meaningful geometrical length scale in the center of the
star where $\Hp$ formally goes to infinity. We use the convection
model by \markcite{1986A&A...160..116K}{Kuhfu{\ss}} (1986) for several reasons.
His model consists of only one equation making it relatively easy to
implement in stellar evolution (\Sec\ref{sec:impOV}),
the equation is derived
from the proper hydrodynamical equations and we trust it to be
physically meaningful for conditions prevalent in
stellar cores. In the stationary, strictly local limit
the cubic equation of MLT is retained when the Ledoux
criterion is employed. And finally, the
\markcite{1986A&A...160..116K}{Kuhfu{\ss}} (1986)-theory gives the same qualitative
behavior for the temperature stratification
in the overshoot region as derived by \markcite{1991A&A...252..179Z}{Zahn} (1991).

\subsection{Implementation of Overshoot in YREC}
\label{sec:impOV}

\subsubsection{Original Treatment}

Core overshoot in YREC has been previously treated
in a way most widely used in stellar evolution codes.
In a first step, the boundary of the convective core
is determined by the Schwarzschild criterion, i.e., regions where
$\nablrad > \nablad$ are labeled as convection zones. Since superadiabaticity
is small in the interior of stars, as discussed before,
the actual stellar temperature gradient
is then set to the adiabatic gradient $\nablad$ within the convection
zone. In a second step, a new boundary is determined by adding a
fraction $\alphOM$ of the pressure scale height to the boundary
at radius $r_\mathrm{s}$ determined by the Schwarzschild criterion
\begin{equation}
r_\mathrm{new} = r_\mathrm{s} + \alphOM\,\Hp(r_\mathrm{s})
\end{equation}
where $\Hp$ is taken at the Schwarzschild boundary.
Finally, all chemical elements are mixed homogeneously within the
new boundary radius given by $r_\mathrm{new}$. By construction,
only the mixed region is extended whereas the
temperature stratification is not affected directly within the
overshoot region. This approach is most correctly termed
\emph{overmixing}.
We label models using overmixing with \emph{CT1}
and call the overshoot parameter in this case $\alphOM$.

\subsubsection{Simple Improvement on Previous Treatment}

It has been pointed out by \markcite{1991A&A...252..179Z}{Zahn} (1991) that the
temperature stratification in the overshoot region is rendered
almost adiabatic due to efficient heat transport in this region and
that this allows for more overshoot than would be possible with
the original stratification. The solutions calculated from
\markcite{1986A&A...160..116K}{Kuhfu{\ss}} (1986)'s model also agree with this assessment.
In order to mimic this effect, we optionally alter the
treatment of overshoot in YREC such that the temperature gradient
is set to the adiabatic gradient also in the overshoot region up to
$r_\mathrm{new}$. In the course of evolution, the convective
core size will differ from the previous approach when adopting the
same overshoot parameter. It is therefore important to make clear
which overshoot prescription was being used when discussing values
of the overshoot parameter. We do so by labeling models using this
approach with \emph{CT2} and denote the overshoot parameter with $\alphOV$.

\subsubsection{Implementation of \markcite{1986A&A...160..116K}{Kuhfu{\ss}} (1986) theory}
The basic equation of \markcite{1986A&A...160..116K}{Kuhfu{\ss}} (1986)'s model is an
equation for the time-development of specific turbulent kinetic energy, $\wturb \, \mbox{[}\mbox{erg\,g}^{-1}\mbox{]}$,
\begin{equation}
  \lderivt{\wturb} - \frac{\Pt}{\rho^{2}}\lderivt{\rho} = \frac{\nablad}{\rho \Hp}\, \jw
  - \frac{c_D}{\Lambda}\, \wturb^{3/2}
  - \gradm{}\left(4 \pi r^2 j_\mathrm{t} \right)
  - \frac{1}{\rho} E_{Q}
\label{eq:wturbeqn}
\end{equation}
where $\mbox{D}/\mbox{D}t$ is the Lagrangian time derivative,
$\rho$ is density, $\Pt$ turbulent pressure, $r$ stellar radius,
$\Hp$ pressure scale height,
$\nablad$ adiabatic gradient, $1/\Lambda=1/(\alphMLT \Hp)+1/(\betr r)$
geometrically limited mixing length scale and $E_{Q}$
viscous energy dissipation.
The first term on the right hand side of this equation is
the driving term for convection and it is proportional to the
convective flux, $\jw$,
\begin{equation}
  \jw = {\alphs}\, \Lambda\, \wturb^{1/2}\,\rho\, T\,
  \frac{\Cp}{\Hp}\,\,\left[ ( \nabla - \nablad )
    - \frac{\varphi}{\delta} \nablmu \right]
  %\left(\gradm{e} - \frac{P}{\rho^2} \gradm{\rho}\right)
\label{eq:jw}
\end{equation}
with
\begin{equation}
  \nablmu     = \frac{d\ln \mu}{d\ln P}
\end{equation}
where $\alphs$ is the turbulent driving parameter,
$\Cp$ the specific heat at constant pressure, and in the
case of an ideal gas with radiation pressure
the dimensional parameters $\delta$ and $\varphi$ take on the
values $\delta = (4-3\beta)/\beta$, $\beta = \Pgas/P$ and
$\varphi=1$ respectively.
The turbulent flux $\jw$
is essentially controlled by the buoyancy term
and the molecular weight gradient in brackets. This term
is positive (negative) in regions where the Ledoux criterion would indicate
convective instability (stability) giving rise to a source (sink)
term in \Eqn(\ref{eq:wturbeqn}) thereby creating (destroying)
turbulent kinetic energy.
Note that the temperature gradient $\nabla$ is not known {\it a priori}.
It is rather a quantity that must be determined from the convection
theory (see Appendix~\ref{apx:algebra}). The second term
on the right hand side of the basic
\Eqn(\ref{eq:wturbeqn}) is a dissipation term with $c_D$ being the
dissipation efficiency parameter and it acts always as a sink. This
term controls the superadiabaticity in the stationary limit of fully
developed convection.

All terms discussed so far are functions
of variables defined locally at each mass shell of the star.
If we were to restrict ourselves to only these variables and assume
time independence, the convective
core size would match exactly the size
determined by the Ledoux criterion or, when neglecting $\nablmu$, by
the Schwarzschild criterion. Indeed, the only deviation from MLT
would be a different superadiabaticity, but since it is
very small in the core it does not make a
difference for the stellar model
whether the temperature gradient $\nabla$
equals $\nablad + 10^{-5}$ or $\nablad + 10^{-6}$.

In view of its relevance to the convective core size the novel
part of \markcite{1986A&A...160..116K}{Kuhfu{\ss}} (1986) theory stems from the
non-local term, the third term in \Eqn(\ref{eq:wturbeqn})
where $\jt$ is given by
\begin{equation}
  \jt = - 4 \pi r^2 \rho^2 \alpht\, \Lambda\, \wturb^{1/2}\, \gradm{\wturb} \punkt
\label{eq:jt}
\end{equation}
Since this term is proportional to the second spatial derivative of
the specific turbulent kinetic energy it leads to overshoot (increase of
convective region beyond Schwarzschild boundary) in regions
of positive curvature and to {\it undershoot}
(decrease of convective region below Schwarzschild boundary)
in regions of negative curvature. In the case of core convection only
the increase of convective core size has been encountered in our studies.
The free parameters $\alphs$ and $c_D$ are factors that
determine the values of convective velocity and convective flux.
As \markcite{1987PhDT.......162K}{Kuhfu{\ss}} (1987) demonstrates, they can be exactly
matched to the ones of MLT by setting:
\begin{eqnarray}
\alphs &=& \frac{1}{2}\sqrt{\frac{2}{3}}\\
 c_D   &=& \frac{8}{3}\sqrt{\frac{2}{3}} \punkt
\end{eqnarray}
We use these standard values throughout this study but 
it must be stressed that this choice is not obligatory and with the
kind of methods described in this paper
it will hopefully prove possible to derive those
parameters from observations as demonstrated here
for the overshoot parameter $\alpht$.

We neglect the second term on the left hand side of the
turbulent kinetic energy equation, since we are
only concerned with the hydrostatic main-sequence phase in which
the star is thermally relaxed and $\mbox{D}\rho/\mbox{D}t$ is close
to zero. For stars that are not in thermal equilibrium, e.g., in the
pre-main sequence phase or in the contraction phase before the onset of
helium burning, this term must be taken into account. We also
neglect the last term in \Eqn(\ref{eq:wturbeqn}), the viscous
energy dissipation. This term is only important for high
velocity flows which are not encountered during the star's main-sequence
lifetime. These simplifications yield
\begin{equation}
  \lderivt{\wturb} = \frac{\nablad}{\rho \Hp}\, \jw
  - \frac{c_D}{\Lambda}\, \wturb^{3/2}
  - \gradm{}\left(4 \pi r^2 j_\mathrm{t} \right) \punkt
\label{eq:wturbeqnsimp}
\end{equation}
We aim at finding a solution for $\wturb$ in the stationary limit
$\mbox{D}\wturb/\mbox{D}t \equiv 0$ and we show in Appendix~\ref{apx:algebra}
that \Eqn(\ref{eq:wturbeqnsimp}) can be cast into the following form:
\begin{eqnarray}
  \lderivt{\wturb} &=&   A\, x_{\wturb} \,{\wturb}^{1/2} - B\,{\wturb}^{3/2} \nonumber\\
                    && + C\, (1-x_{\wturb}) \gradn{\wturb^{3/2}} + D\, \gradn{}\!\left(E\, \gradn{{\wturb}^{3/2}}\right) \nonumber \\
  \label{eq:wturbeqmath}
\end{eqnarray}
with
\begin{equation}
x_{\wturb} = \frac{1}{1 + F\, \wturb^{1/2}}
\end{equation}
where $A$--$F$ are functions of the known stellar background model.
Without the non-local terms, finding a stationary solution is simple,
similar to MLT, one has to solve a cubic algebraic equation
at every mass shell. However, with the spatial coupling an analytic
expression is not obvious. Therefore, we solve the non-linear,
time-dependent \Eqn(\ref{eq:wturbeqmath}). There is
just one further complication. In radiative regions 
\Eqn(\ref{eq:wturbeqmath}) does not have a stationary solution.
%the turbulent kinetic energy is decaying exponentially which
%is physically reasonable. 
One might be tempted to
set $\wturb = 0$ in those regions which is a trivial solution to
\Eqn(\ref{eq:wturbeqmath}). However, the zero solution branch is
completely separated from the finite solution path and therefore
$\wturb$ will always remain zero. In reality, some finite
turbulent kinetic energy is always maintained due to thermal noise.
Technically, this problem can be overcome by adding an artificial term
to the right hand side of \Eqn(\ref{eq:wturbeqmath}):
\begin{equation}
+ \frac{\left| A \right|}{\wturb}\,(10^{-10})^{3/2} \punkt
\end{equation}
This artificial term can be motivated based
on the physical picture that it maintains a noise, or seed convection
underground of $\wturb = 10^{-10}$ in radiative regions.
It must be strongly emphasized, that seed
values about two orders of magnitude higher start to falsify the solution.
Order of magnitude lower values are possible in principle but
one quickly runs into problems related to lack of numerical precision.
With this carefully chosen artificial term, we first initialize
$\wturb$ to some arbitrary value (typically $10^{-10}$) and subsequently
use an implicit time integration method to reach the equilibrium state. We
utilize LIMEX (Linear IMplicit EXtrapolation Method) \markcite{Ehrig2003}({Ehrig} \& {Nowak} 2002)
for performing the time-integration. The solution yields the
specific turbulent kinetic energy $\wturb$ at every mass shell. We define
shells to be convective, if:
\begin{equation}
  x_{\wturb} < 0.1
\end{equation}
thereby determining the size of the convective core. The boundary of
the convective region is sharply defined by an extremely steep
falloff of $\wturb$. In fact, the transition region can be
estimated and is on the order of $100\,\km$, orders of magnitudes
below the model resolution. Finally, the
temperature gradient can be calculated from (see Appendix~\ref{apx:algebra}):
\begin{equation}
\nabla = \nablad + x_{\wturb}\, (\nablrad-\nablad) + (1-x_{\wturb})
         \, \left( G \gradn{\wturb} + H \right)
\end{equation}
and it remains close to the adiabatic one over the complete
convection region.

We refer to models that were calculated with the Kuhfu{\ss} convection
theory with \emph{CT3} and the associated overshoot parameter
with $\alpht$.

\subsection{Kuhfu{\ss}- versus traditional Overshoot}

A first comparison of core sizes between MLT and KCT has been
given by Kuhfu{\ss} himself in his PhD thesis \markcite{1987PhDT.......162K}({Kuhfu{\ss}} 1987).
For a $15\,\Msun$ with $\alphMLT=1.5$ and $\alpht=0.25$ he deduces
a core overshoot parameter equivalent to
$0.4\,\Hp$ above the MLT Schwarzschild boundary.
Furthermore, his analysis gives large overshoot of $0.3\,\Hp$ even
for the lowest mass stars around $1.1\,\Msun$.

However, as has been pointed out by \markcite{1998A&A...340..419W}{Wuchterl} \& {Feuchtinger} (1998) in the
context of the KCT when applied to RR-Lyrae stars, the
pressure scale height $\Hp$ must be limited where it exceeds the
geometrically possible length scale. In the case of RR-Lyrae stars,
the pressure scale height is limited by the distance to the stellar
surface.

A similar situation occurs in the deep interior of stars. The
pressure scale height goes to infinity when approaching the
stellar center rendering it an inadequate measure of scale length
in regions close to the center. Again, it is physically plausible
to limit the length scale to the geometrical distance of the
center yielding a length scale $\Lambda$ of:
\begin{equation}
\Lambda = \left( \frac{1}{\alphMLT \Hp} + \frac{1}{\betr r} \right)^{-1}
\end{equation}
introducing the parameter $\betr$ which we set to $1.0$ throughout this
whole paper. In a non-local theory of convection this limited
pressure scale height in the central regions influences not only the core
size of stars with small convective cores but it is also important
for all stellar core sizes and masses.

\begin{figure}
\includegraphics[height=\textwidth,angle=-90]{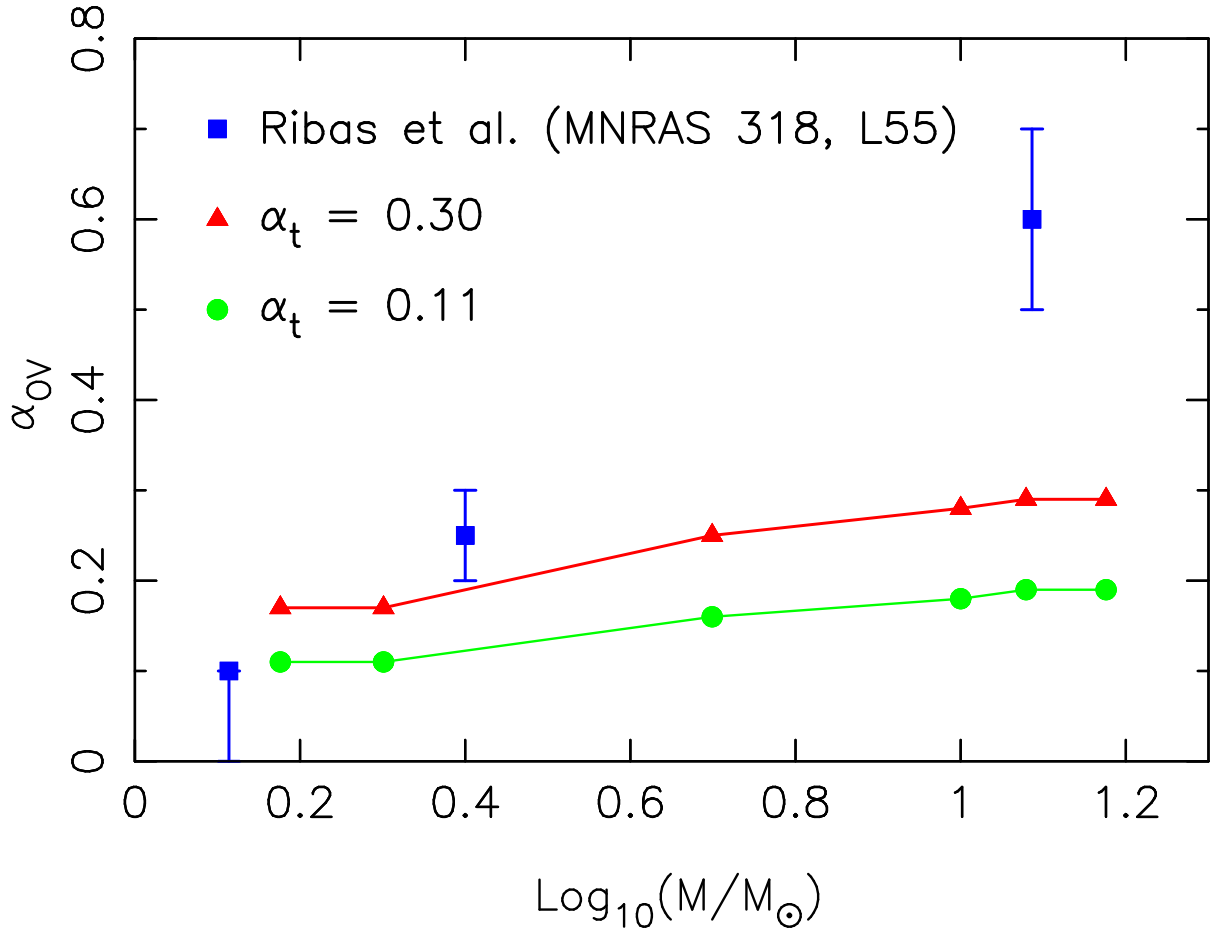}
\caption{Core overshoot parameter measured
in fractions of the pressure scale
height for KCT with $\alpht=0.11$ (circles)
and $\alpht=0.30$ (triangles) for stellar masses
between $1.5\,\Msun$ and $15.0\,\Msun$. The overshoot from detached
eclipsing binaries derived by Ribas et al.\ is given for
comparison. KCT supports a moderate increase in
overshoot with increasing stellar mass. Overshoot of $0.6$
as claimed by Ribas et al. at the high mass end cannot be
explained with KCT.\label{fig:ribasvgl}}
\end{figure}

With this modified length scale, we repeat the comparison
between MLT and KCT for two cases, one with $\alpht=0.11$
and the other with $\alpht=0.30$ both values in accord with
stellar evolution of \ProcA (see \Sec\ref{sec:constrOV}). We
determine the amount of overshoot -- measured in pressure
scale heights -- needed to reproduce the models employing
KCT.  As can be seen in
\Fig\ref{fig:ribasvgl}\notetoeditor{Please print ALL Figures in color ONLY
in electronic version.} our models show less overshoot
compared to Kuhfu{\ss}'s original analysis: with
$\alpht=0.3$ we find $\alphOV=0.17$ for a $1.5\,\Msun$ star
and $\alphOV=0.29$ for a $15\,\Msun$ star. Since
\markcite{2000MNRAS.318L..55R}{Ribas} {et~al.} (2000) claim an increase of core
overshoot for increasing mass by measurements of eclipsing
binaries we also ask whether KCT can naturally explain this
behavior. Indeed, we can see a moderate increase of
overshoot with mass. Nonetheless, the increase is not
pronounced enough to explain the large overshoot of $0.6$
claimed for the $11.1\,\Msun$ star {\sl V380~Cyg}.  By
increasing the mixing length parameter $\alphMLT$ the
increase of overshoot with mass can be enhanced slightly but
an overshoot of $0.6$ remains out of reach.

It is obvious to think rotation could play an important role in
increasing the core through rotational mixing in a star of such high
mass. \markcite{2000ApJ...544..409G}{Guinan} {et~al.} (2000) have carried
out a detailed analysis of this system reaching the conclusion that
the slow rotation of this star cannot account for substantial
additional core size increase. 
The recently derived overshoot of $\alphOV=0.10\pm0.05$ for another
high mass star with $9\,\Msun$
\markcite{2003Sci...300.1926A,2004A&A...415..251D}({Aerts} {et~al.} 2003; {Dupret} {et~al.} 2004)
seems to indicate that the large overshoot for
{\sl V380~Cyg} may not be a universal property of high mass stars.

\section{Stellar Models of \ProcA}

\subsection{Constraints from Observations}
\label{sec:OBSconstraints}
\ProcA and its companion white dwarf constitute a visual binary system
with a 40 yr period. The redetermined mass of \ProcA based on
250 photographic plates of observations between 1912 and 1995
combined with modern direct measurements of the angular
separation of the pair utilizing  HST Planetary Camera and
ground based coronagraph data are given by \markcite{2000AJ....119.2428G}{Girard} {et~al.} (2000)
with a derived mass of $1.497 \pm 0.037\,\Msun$.
The same study also determines the parallax of \ProcA to be
$0\farcs2832 \pm 0\farcs0015$ a value different from the {\it Hipparcos}
result $p = 0\farcs28593 \pm 0\farcs00088$. We take the mean value of
both independent measurements and adopt the difference to the upper and
lower bound of these measurements as our $1\sigma$ error estimate
thereby yielding $p = 0\farcs28457 \pm 0\farcs0025$. The larger adopted
uncertainties for the parallax lead to larger error bars for
\ProcAs radius and luminosity than otherwise expected if we had used
the quoted errors from the actual measurements.

The radius of \ProcA can be retrieved using the stellar
angular diameter data derived from optical interferometry
$\theta = (5\farcs51 \pm 0\farcs05) \times 10^{-3}$
\markcite{1991AJ....101.2207M}({Mozurkewich} {et~al.} 1991). Taking the center-to-limb variation
from detailed model atmospheres into account \markcite{2002ApJ...567..544A}{Allende Prieto} {et~al.} (2002)
correct this value slightly:
$\theta = (5\farcs48 \pm 0\farcs05) \times 10^{-3}$.
When combined with parallax of $p = 0\farcs28457 \pm 0\farcs0025$ and
applying the laws of error propagation to the three quantities involved
one finds the radius
\begin{equation}
R \simeq \frac{\theta}{2p\, \tan(\theta_{\sun}/2)}
= 2.070 \pm 0.026 \,\Rsun
\end{equation}
adopting $\theta_{\sun}/2=959\farcs64 \pm 0\farcs02$
\markcite{1999A&AS..139..219C}({Chollet} \& {Sinceac} 1999). This result is statistically identical to
the independent angular diameter measurement of
\markcite{2004A&A...413..251K}{Kervella} {et~al.} (2004).

The effective temperature is measured with the
averaged bolometric flux \markcite{1997A&A...323..909F}({Fuhrmann} {et~al.} 1997)
$F_\mathrm{BOL} = (18.20 \pm 0.43) \times 10^{-6}\,\, \mbox{erg}\,\mbox{cm}^{-2}\,\mbox{s}^{-1}$:
\begin{equation}
\Teff = 7400 \times \left(\frac{F_\mathrm{BOL}}{\theta^2}\right)^{\!\!{1/4}}
= 6530 \pm 50 \,\mbox{K}
\end{equation}

With an absolute distance of \ProcA from earth given by the
parallax measurement, $d=3.0857\times 10^{18}\,\mbox{cm}/p(\arcsec)$
we derive the intrinsic luminosity
($\Lsun = 3.833 \times 10^{33}\,\,\mbox{erg}\,\mbox{s}^{-1}$)
\begin{equation}
L = 4 \pi d^2 F_\mathrm{BOL} = (7.016 \pm 0.21) \,\Lsun
\end{equation}
and we use the value $\log_{10}(L/\Lsun) = 0.8461 \pm 0.013$
throughout this study.

\subsection{Constructing Tracks consistent with Observations}
\label{sec:consistOBS}
\begin{figure}
\includegraphics[height=\textwidth,angle=-90]{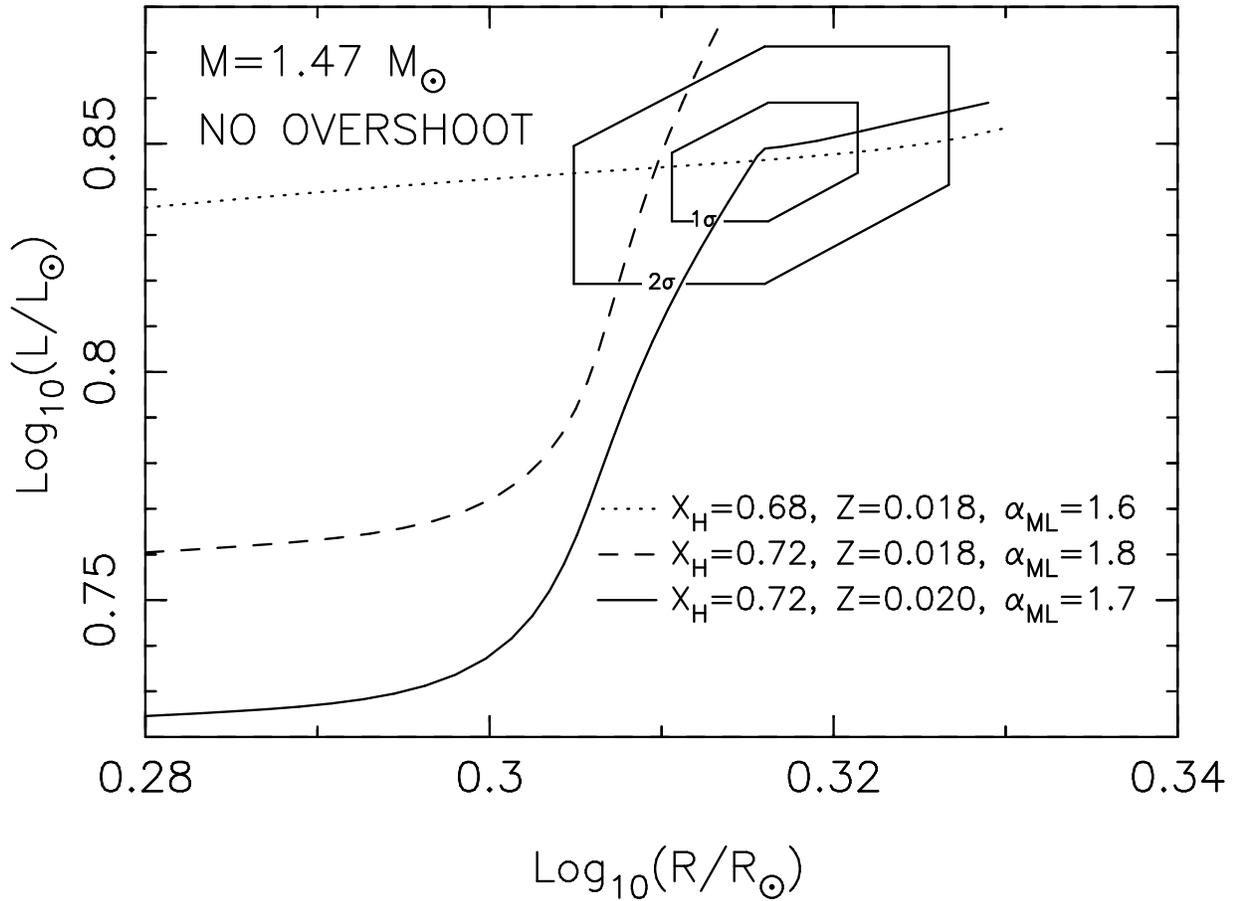}
\caption{Models with no overshoot are in accord with
observational constraints. Models are found in either
the (late) main-sequence (dotted, dashed lines) or the subgiant phase
(solid line). Subgiant models can only be found for
initial masses less than $1.47\,\Msun$ and are completely
excluded when tiny amounts of overshoot of $\alphOV > 0.05$ are
assumed. The \emph{error box} is given by the observational
constraints on luminosity, radius and effective temperature.
Its hexagonal shape arises from making use of all
three constraints.\label{fig:noov}}
\end{figure}

\begin{figure}
\includegraphics[height=\textwidth,angle=-90]{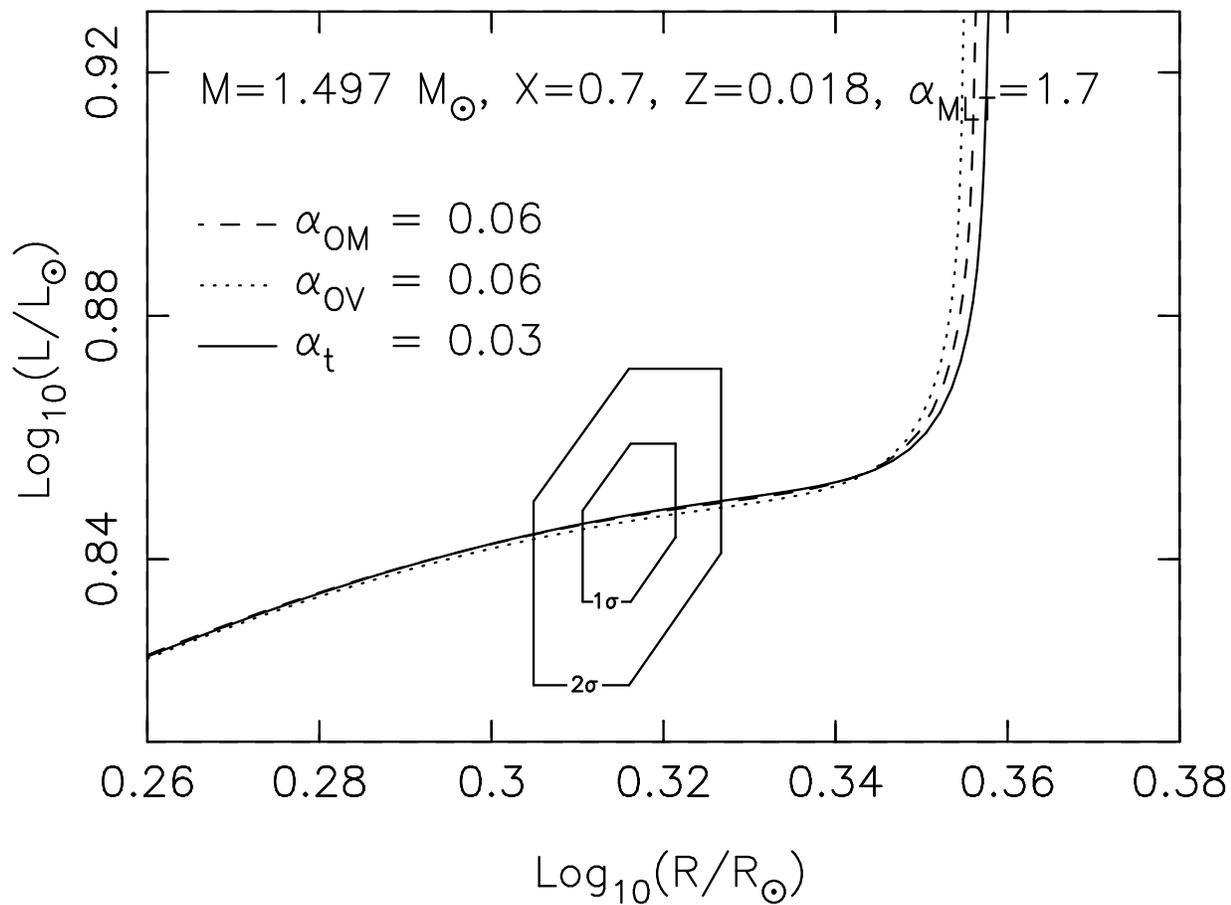}
\caption{Best fit overshoot parameters for three
different overshoot descriptions. The best fit values
indicate only small overshoot but due to the large
uncertainties overshoot could range from $\alphOM=0\ldots1.18$.
\label{fig:bestfit}}
\end{figure}

Since parallax $p$, angular diameter $\theta$ and bolometric flux
$F_\mathrm{BOL}$ are measured with completely independent
techniques and data sets
the derived basic stellar parameters luminosity
$L$, radius $R$ and effective temperature $\Teff$ can be jointly used
to constrain stellar models for \ProcA. Therefore
we require all stellar models for \ProcA to fall into a
three dimensional {\it error box} given by observational constraints
for $L$, $R$ and $\Teff$. With help of the
identity $L=4 \pi R^2 \sigma \Teff^{4}$ one of the constraints
can be projected onto a plane given by the remaining two. If
we look, for example, at the $R$-$L$-plane 
the error box is given by a hexagonal shape.
Both $1\sigma$ and $2\sigma$ limits as inferred from
\Sec\ref{sec:OBSconstraints} are shown
(\Fig\ref{fig:noov}\notetoeditor{\Fig\ref{fig:noov} and
\Fig\ref{fig:bestfit} may appear side-by-side.}).

Our model calculations start from the ZAMS with the default
parameters of mass $M=1.497\,\Msun$, solar hydrogen content $X = 0.70$,
solar metallicity $Z=0.018$, mixing length $\alphMLT=1.7$,
no overshoot and no envelope element diffusion.
%Note that a model with these default values does not
%fall into the error box and therefore does not qualify as a model
%for \ProcA.
We now vary the mass
$M=\{1.423$, $1.435$, $1.45$, $[1.46$, $1.47$, $1.48$, $1.485$, $1.49$,
$1.497$, $1.509$, $1.515$, $1.52$, $1.53$, $1.534]$, $1.55$, $1.56$, $1.571\}$,
hydrogen content $X=\{0.68$, $[0.69$, $0.70$, $0.71]$, $0.72\}$,
metallicity $Z=\{0.016$, $[0.017$, $0.018$, $0.019]$, $0.020\}$,
mixing length parameter $\{[1.5$, $1.6$, $1.7]\}$ and
convection description $CT=\{[1$, $2$, $3]\}$ with the
associated overshoot parameters
$\alphOM=\{[0.0$, $0.2$, $0.4$, $0.6$, $0.8]\}$,
$\alphOV=\{[0.2$, $0.4$, $0.6]\}$ and
$\alpht=\{[0.2$, $0.6$, $3.0]\}$.
The full set of all combinations required us to calculate $14025$ 
different models. Of these, $2938$ fell in the observational
%2144(CT1+CT2)+794(CT3)
$2\sigma$ error box and $1541$
%1147(CT1+CT2)+394(CT3)
in the $1\sigma$ error box.
In order to explore the sensitivity of our results to uncertainties
in the input parameters, we also study
a more {\it restricted} parameter set given by all combinations
within the values of the innermost brackets.
The restricted set consists of $3564$ models, of which $712$
%438(CT1+CT2)+274(CT3)
qualify within the $2\sigma$ and $354$
%208(CT1+CT2)+146(CT3)
in the $1\sigma$ observational limits. Models with overshoot
are calculated in the case of three different overshoot prescriptions
\emph{CT1-CT3} by varying the overshoot parameter relevant to the adopted
theory.
%A variant of the KCT but with the
%$\mu$-gradient terms neglected is studied and named \emph{CT4}.
%We also calculate two additional models
%to see the effects of element diffusion on the model.

For each stellar track that falls within the error box we pick
out five models, two at the inner and outer $2\sigma$ boundaries,
two at the inner and outer $1\sigma$ boundaries and finally
one close to the center of the error box. In instances
in which models do not cross the $1\sigma$ box only two models are
picked out. On all picked out models, a full pulsational 
analysis is performed and discussed in \Sec\ref{sec:pulseana}.

All models within the $1\sigma$ error box are found
to be in the main sequence phase of evolution.
\markcite{1999ApJ...525L..41C}{Chaboyer}, {Demarque}, \&  {Guenther} (1999) also consider a model in the
hydrogen shell burning subgiant phase. If we allow the full $2\sigma$
we also cannot exclude shell burning phase models from stellar
tracks. However, subgiant models are only found for
initial masses of $M \le 1.47\,\Msun$.
It is worth noting, that with overshoot, it
becomes increasingly harder to bring a shell burning model
in accord with the observational
($2\sigma$) constraints. By calculating two
additional sets of models with $\alphOM = 0.02$ and $\alphOM = 0.05$
and carefully taking into account all possible combinations within the
remaining parameters we could verify that tiny amounts of overshoot of
$\alphOM > 0.05$ completely inhibit the shell burning model
as a viable possibility.

\subsection{Constraints on Core Overshoot}
\label{sec:constrOV}
As can be seen in \Fig\ref{fig:noov} models with
{\it no} overshoot are consistent with the
observational constraints. At present,
we cannot claim evidence for or rule out overshoot on the basis
of traditional stellar modeling.  But we can try to
provide upper limits and best fit values
on overshoot for the different overshoot prescriptions.

Since the stellar luminosity is --- to first order --- a function of
mass, a lower mass value for \ProcA enforces more overshoot in an effort
to increase the model luminosity to the observed value. Overshoot
generally causes larger convective cores, thereby transforming
more hydrogen into helium which in the course of
nuclear burning leads to a higher mass averaged molecular weight
$\bar{\mu}$. Being also proportional to $\bar{\mu}^4$ the
luminosity is sensitively increased by small amounts of overshoot.

Best fit values are ascertained by adopting default stellar parameters
and adding overshoot until the tracks coincide with the best fit
locus defined halfway between the radius  and
halfway between the luminosity uncertainties, the latter taken
at the best fit radius. The best fit values
are $\alphOM=0.06$, $\alphOV=0.06$ and $\alpht=0.03$
(\Fig\ref{fig:bestfit}).

Smaller mass and $\alphMLT$ and larger $X$, $Z$ shift tracks to
lower luminosities thereby requiring more overshoot in order to
bring the tracks back into the observational error box. We
now simultaneously alter all four input parameters such that
maximal overshoot is needed.

\begin{figure}
\includegraphics[height=\textwidth,angle=-90]{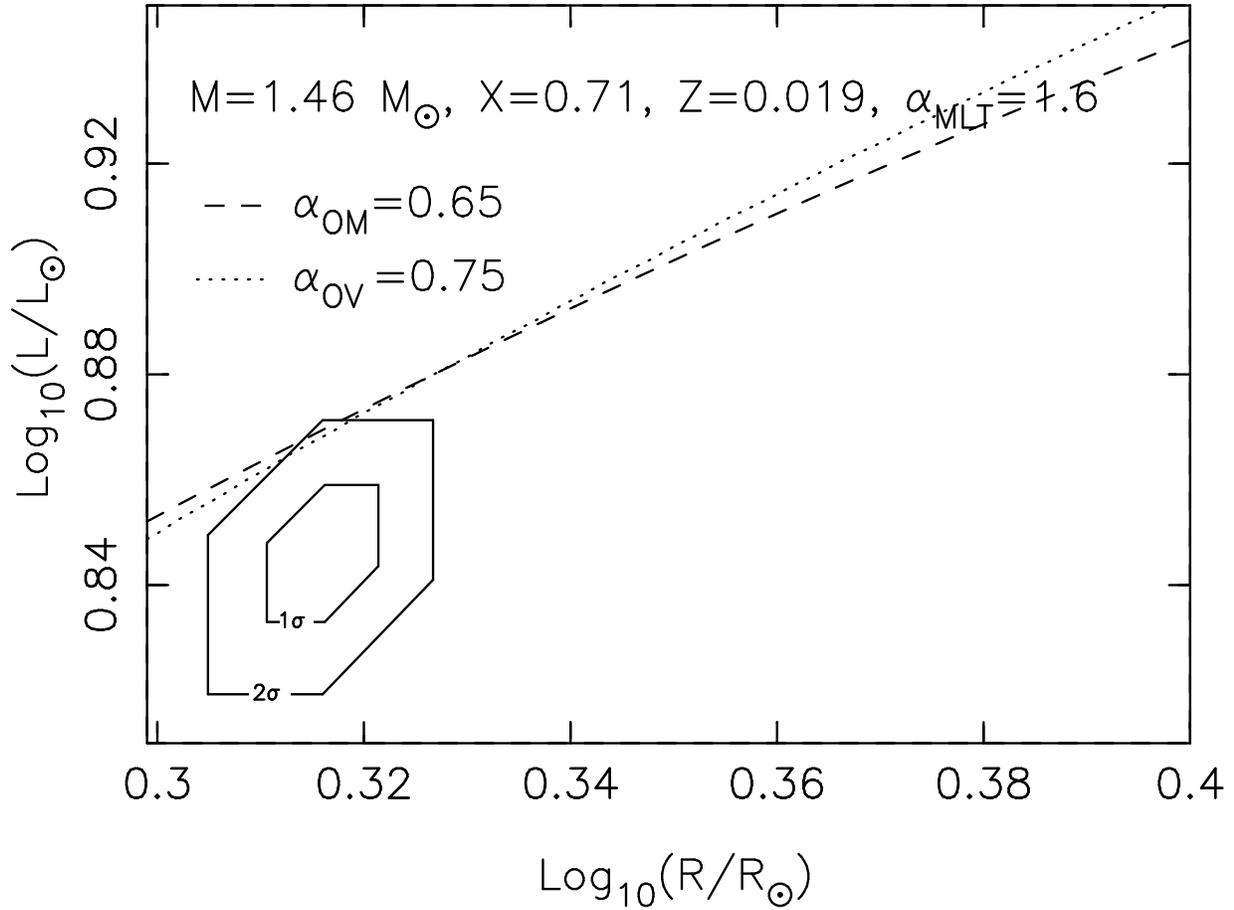}
\caption{Maximally allowed overshoot parameters for
different overshoot descriptions when adopting the
restricted input parameter set (\emph{weak limits})
but still allowing for the full $2\sigma$ observational
error box. The model using KCT cannot be shifted into the
error box despite extremely large values for $\alpht$.
\label{fig:weaklimits}}
\end{figure}

\begin{figure}
\includegraphics[height=\textwidth,angle=-90]{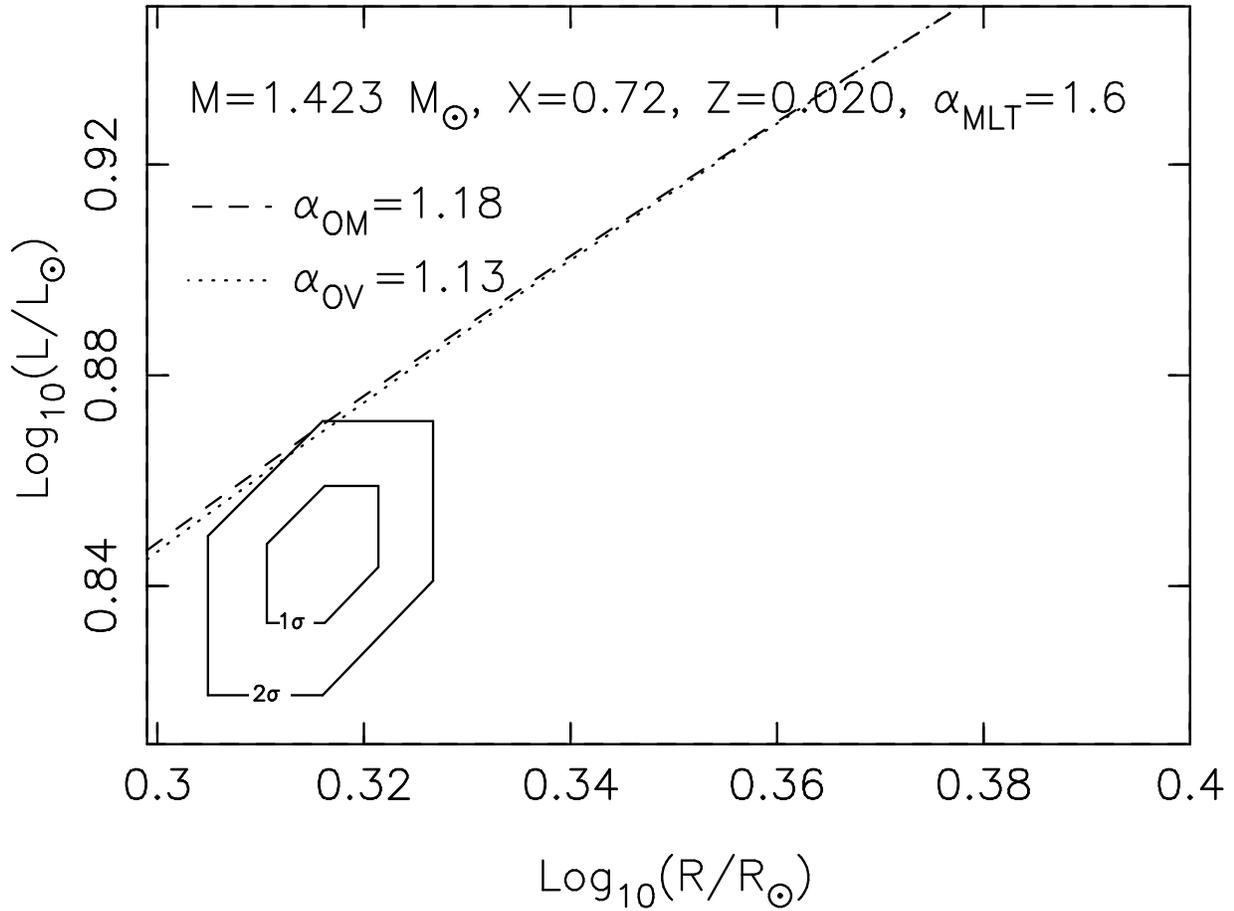}
\caption{Maximally allowed overshoot parameters for
different overshoot descriptions when adopting the
full input parameter set (\emph{strong limits}).
The model using KCT cannot be shifted into the
error box despite extremely large values for $\alpht$.
\label{fig:stronglimits}}
\end{figure}

The first set of upper overshoot limits is derived by
using a mass of $M=1.46\,\Msun$, $\alphMLT=1.6$ and
hydrogen $X=0.71$ and metallicity of $Z=0.019$. These input
parameters correspond to the set of restricted limits
\Sec\ref{sec:consistOBS}.
We denote the limits by this procedure {\it weak limits} and
they are $\alphOM \le 0.65$, $\alphOV \le 0.75$.
(see \Fig\ref{fig:weaklimits}\notetoeditor{\Fig\ref{fig:weaklimits} and
\Fig\ref{fig:stronglimits} may appear side-by-side.}).

The second set of upper limits, the {\it strong limits}, are found
by repeating this procedure for the minimally allowed
observational mass of $M=1.423\,\Msun$, 
$\alphMLT=1.6$ and the highest values allowed for in this
study for hydrogen and the metallicity, i.e., $X=0.72$ and
$Z=0.020$.  Thus we take the most extreme
values in favor of overshoot.
The strong limits are $\alphOM \le 1.18$ and $\alphOV \le 1.13$
(see \Fig\ref{fig:stronglimits}).

An interesting property of the \markcite{1986A&A...160..116K}{Kuhfu{\ss}} (1986)-theory
is, that the core size cannot be increased above a certain limit
regardless of the overshoot parameter $\alpht$. For Procyon,
the extent of the overshoot region is not substantially increased
above $\alpht=5$. This behavior is rooted in the KCT and we
will discuss this in more detail in the next section
(\Sec\ref{sec:extKCT}).

\subsection{Extent of Overshoot region with KCT}
\label{sec:extKCT}
It is important to point out that the role of the free parameter
$\alpht$ is different compared to the free parameters $\alphOM$
or $\alphOV$. Increasing the latter two will always lead to more
extended convective cores --- up until the star is fully convective.
On the other hand, $\alpht$ is a factor regulating the relative
importance of the non-local terms, i.e., the third term in
\Eqn~\ref{eq:wturbeqnsimp}. Increasing the non-local coupling does
not always lead to more extended overshoot. In fact, extremely large
values for $\alphOV$ lead to smaller overshoot extents.

\begin{figure}
\includegraphics[height=0.49\textwidth,angle=-90]{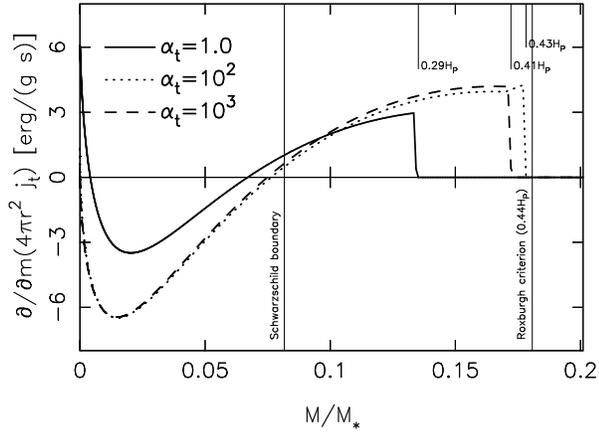}
\includegraphics[height=0.49\textwidth,angle=-90]{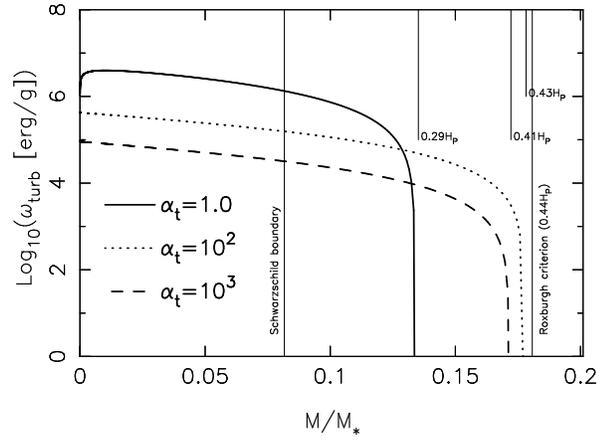}
\caption{
Run of non-local convection term of equation (\ref{eq:wturbeqn})
for KCT (left panel) and the corresponding turbulent kinetic energy
(right panel). Increasing the overshoot parameter $\alpht$ not
necessarily increases the overshoot extent.
\label{fig:oversh}}
\end{figure}

The run of the non-local term is shown
in \Fig~\ref{fig:oversh} (left panel)\notetoeditor{Two panels in
\Fig\ref{fig:oversh} must appear side-by-side.}
for three different and extremely high
values for $\alpht$. The largest convective cores can be
attained with $\alpht \sim 100$ which corresponds to
an overshoot distance of $\sim 0.4\,\Hp$.
The model with $\alpht = 1000$ produces smaller overshoot regions.
The reason for this lies in the run of the non-local term itself:
above the Schwarzschild boundary this term is positive and acts towards
more extended cores but in the inner regions it is negative
and acts to decrease the overshoot distance. The net effect can be studied
in \Fig~\ref{fig:oversh} (right panel) where we show the corresponding
stationary solution of the turbulent kinetic energy. High values
of $\alpht$ finally decrease the turbulent kinetic energy in the
inner part and thereby the velocity of the eddies at the
Schwarzschild boundary such that they lack the energy to travel far.

In both panels of \Fig~\ref{fig:oversh}
we show how the maximum core size derived from
the Kuhfu{\ss} theory compares with the well known
integral criterion derived by \markcite{1978A&A....65..281R}{Roxburgh} (1978). As
can be seen from the figures, the maximally allowed sizes by
KCT almost exactly match the core size predicted by the
integral criterion.

With this property of the more physically motivated overshoot
prescription it would be possible to constrain the uncertainties
of mass, hydrogen content, metallicity and mixing-length parameter.
Since the scope of this paper is to find {\it observational} constraints
from traditional stellar modeling when combined with seismological
data we do not explore this possibility in more detail. However,
when comparing the different overshoot prescriptions in the next
sections we must keep in
mind that models of \ProcA employing the Kuhfu{\ss}-theory
inherit convective cores with an overshoot distance
not larger than $\sim 0.4\,\Hp$.

\section{Pulsation Analysis of \ProcA}
\label{sec:pulseana}
\subsection{p-modes}
\label{sec:pmodes}
\begin{figure}
\includegraphics[height=\textwidth,angle=-90]{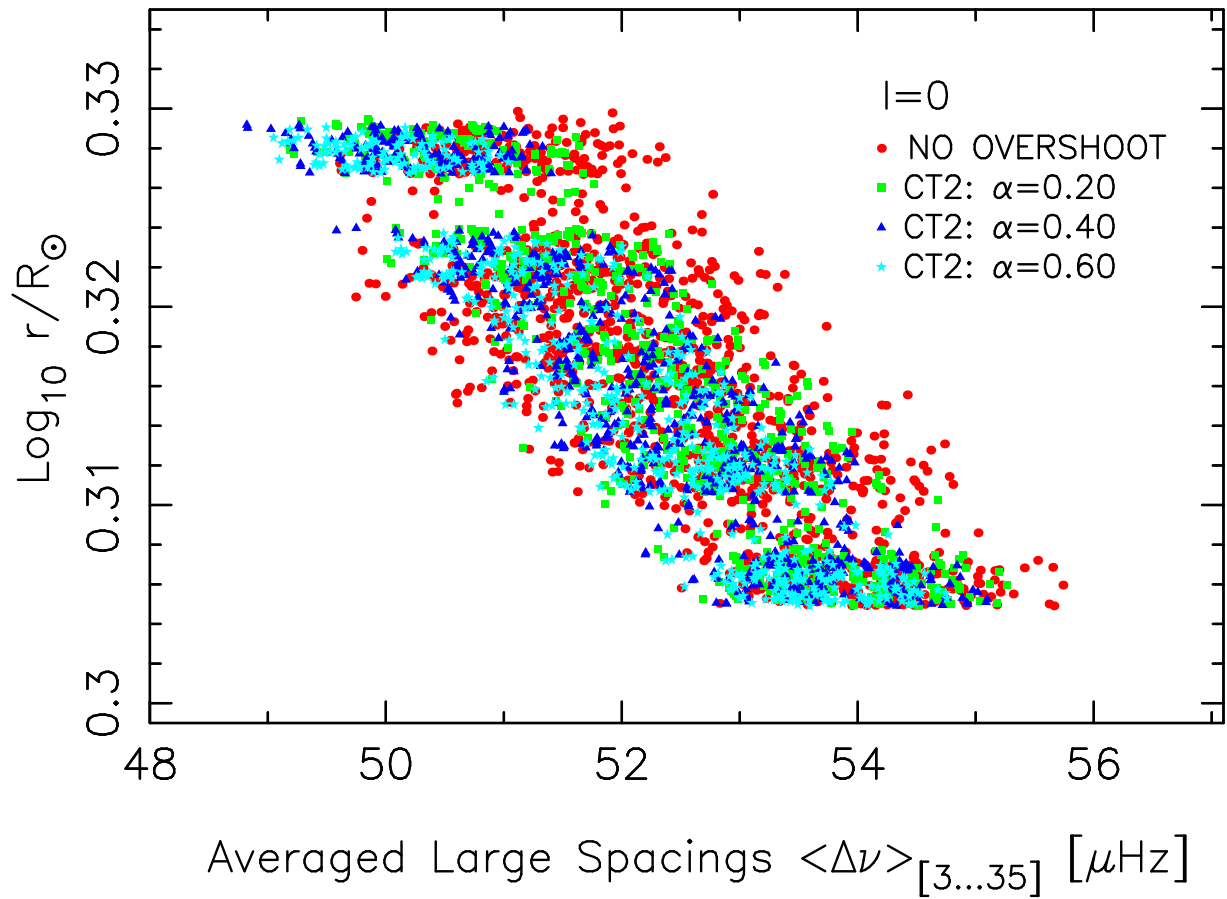}
\caption{The averaged large spacings are sensitive to the
stellar radius as expected. They can be used to measure
the star's radius more precisely.\label{fig:pm_large}}
\end{figure}

Ground based attempts to measure the p-mode large spacings
have so far been inconclusive
\markcite{1991ApJ...368..599B,1995MNRAS.273..367B,
1998A&A...340..457M,1999A&A...350..617B,1999A&A...351..993M}({Brown} {et~al.} 1991; {Bedford} {et~al.} 1995; {Mosser} {et~al.} 1998; {Barban} {et~al.} 1999; {Marti{\' c}} {et~al.} 1999).
Of these works \markcite{1998A&A...340..457M}{Mosser} {et~al.} (1998) derive a tentative
value for the large spacing of $53 \,\mu\mbox{Hz}$.
\markcite{1999A&A...350..617B}{Barban} {et~al.} (1999) and \markcite{1999A&A...351..993M}{Marti{\' c}} {et~al.} (1999)
claim to have observed excess power and spacings of $55 \,\mu\mbox{Hz}$
which would be
compatible with the theoretical predictions (see \Fig\ref{fig:pm_large}).
However, all these measurements have been restricted
to single site measurements producing data contaminated by significant
amplitudes of daily aliases.

While p-mode frequencies as a function of radial and
azimuthal order $\nu(n,l)$
hold the most accurate and detailed information about
the run of sound speed within a star, one cannot hope to match
the model to all frequencies by blindly searching the allowed
parameter space. This
procedure would only work if a stellar model matched the
star as well as our solar models match the Sun. Also,
inversion techniques cannot be easily applied to stars
because we are only able to observe low azimuthal order
p-modes, hence, cannot construct detailed enough kernels to
unambiguously span the surface layers.
Nonetheless, we can make use of averaged quantities which
dispense some of the rich information contained in the oscillation
spectrum but still pose additional and useful constraints on
the stellar model. Such information is available from the \emph{averaged}
first- and second-order frequency spacings
\markcite{1984srps.conf...11C}(cf., {Christensen-Dalsgaard} 1984, p.~11).

The frequency spacings for \ProcA have been calculated and
analyzed extensively before
\markcite{1993ApJ...405..298G,1999ApJ...525L..41C}({Guenther} \& {Demarque} 1993; {Chaboyer} {et~al.} 1999).
They noted that mode bumping occurs in \ProcA and that the
irregular spacing associated with mode bumping could be
used to pin down Procyon's evolutionary phase. Independent
of the oscillation data, they also noted an inconsistency
between the astrometric mass of Procyon and the mass
obtained from fitting their models to Procyon's position
in the HRD, with the latter turning out to be correct
\markcite{2000AJ....119.2428G}({Girard} {et~al.} 2000).

In this paper we discuss what existing observational data
can tell us about convective
core overshoot, and we ask how future observational data,
including seismic data, can help constrain the models further.

For all our models that fall within the error box we calculate the
$\lz,1,2,3$ p-modes from radial order $n = 0$ up to about $35$ 
using the non-radial pulsation code JIG developed by
\markcite{1994ApJ...422..400G}{Guenther} (1994). Mode bumping for \ProcA occurs
at low radial orders between $\sim 0$ -- $16$. Mode bumping, also
referred to as avoided crossings, occurs in more evolved
stars when the molecular weight gradient left behind by
nuclear burning leaves a barrier that traps g-modes in the
core. When the frequencies of the g-modes increase, as the
star evolves, they eventually cross through the frequencies
of the lowest radial order p-modes. In regions of frequency
overlap the modes mix, perturbing the frequencies.
Therefore, mixed modes contain extremely interesting information
on core overshoot. Since we are dealing with
a resonance phenomenon very accurate stellar models are required.
One of us (D.B.G) is exploring these possibilities in a subsequent
paper.

The p-modes with radial order greater than about $16$ show
regular spacings in all our models. It is this information that
is likely the first to be detected and measured accurately.

\subsubsection{Large Spacings}
\label{sec:largefreq}
The first order frequency spacings or large spacings are defined as
\begin{equation}
\Delnunl = \nu(n,l) - \nu(n-1,l) \punkt
\end{equation}
It follows from \markcite{1980ApJS...43..469T}{Tassoul} (1980) asymptotic theory that
high radial order modes are approximately equally spaced with
\begin{equation}
\Delta \nu \simeq \left( 2 \int_0^R \frac{dr}{c} \right)^{-1}
\end{equation}
where $R$ is the radius and $c$ is the sound speed.
Since the integrand is proportional to the inverse of the
sound speed the large spacings are most sensitive to the outer
layers of the star where $c$ is smallest. The
spacings depend on the radius and the mass of the star with 
$\delta \nu/\nu \propto \delta r/R_{*} - \delta M/(3M_{*})$.
Can we improve on \ProcAs radius estimate
by making use of the information provided by the large spacings?
Indeed we can, as seen in \Fig\ref{fig:pm_large}. Here we plot
the large spacings averaged over the range $n=[3\ldots35]$
versus stellar radius for the order $\lz$. 
Each point in the
diagram corresponds to a calculated stellar model with $CT2$.
The spread from a
straight line shows the full set of model uncertainties due to different
mass, overshoot, metallicity and mixing length parameters. We verified
that models corresponding to the other convection prescriptions, i.e.,
from $CT1$ and $CT3$ do not widen the spread. The spread is
tight enough to improve on the radius by a factor of two
for measured averaged large spacings $\lesssim 51\,\mu\mbox{Hz}$
and $\gtrsim 54\,\mu\mbox{Hz}$. The radius  measurement in the
range $51$-$54\,\mu\mbox{Hz}$ can be improved upon only with smaller
improvement factors and it is impossible to do so in the range
$52$-$52.5\,\mu\mbox{Hz}$. The $l=1,2$ modes could provide additional
independent measurements but it turns out that there is more than one
region where models cluster which leads to a wider spread.

Since the large spacings are sensitive to the outer layers we do not
expect them to be sensitive to different core overshoot. For
different orders of $l$ we looked at the convective core size
as a function of the averaged large spacings.
The points scatter all over the diagram
so no correlation between the large spacings and the convective
core size exists. This fact is indeed fortunate if we want to use
the large spacings for an independent measure of the stellar radius.

\subsubsection{Small Spacings}
The small spacings are defined by
\begin{equation}
\delnunl = \nu(n,l) - \nu(n-1,l+2) \punkt
\end{equation}
Again, asymptotic approximations predict equal spacings
proportional to an integral with an integrand of $r^{-1}(dc/dr)$
which increases with depth. Thus, the small spacings are most
sensitive to the interior where the integrand is largest.

\begin{figure}
\includegraphics[height=\textwidth,angle=-90]{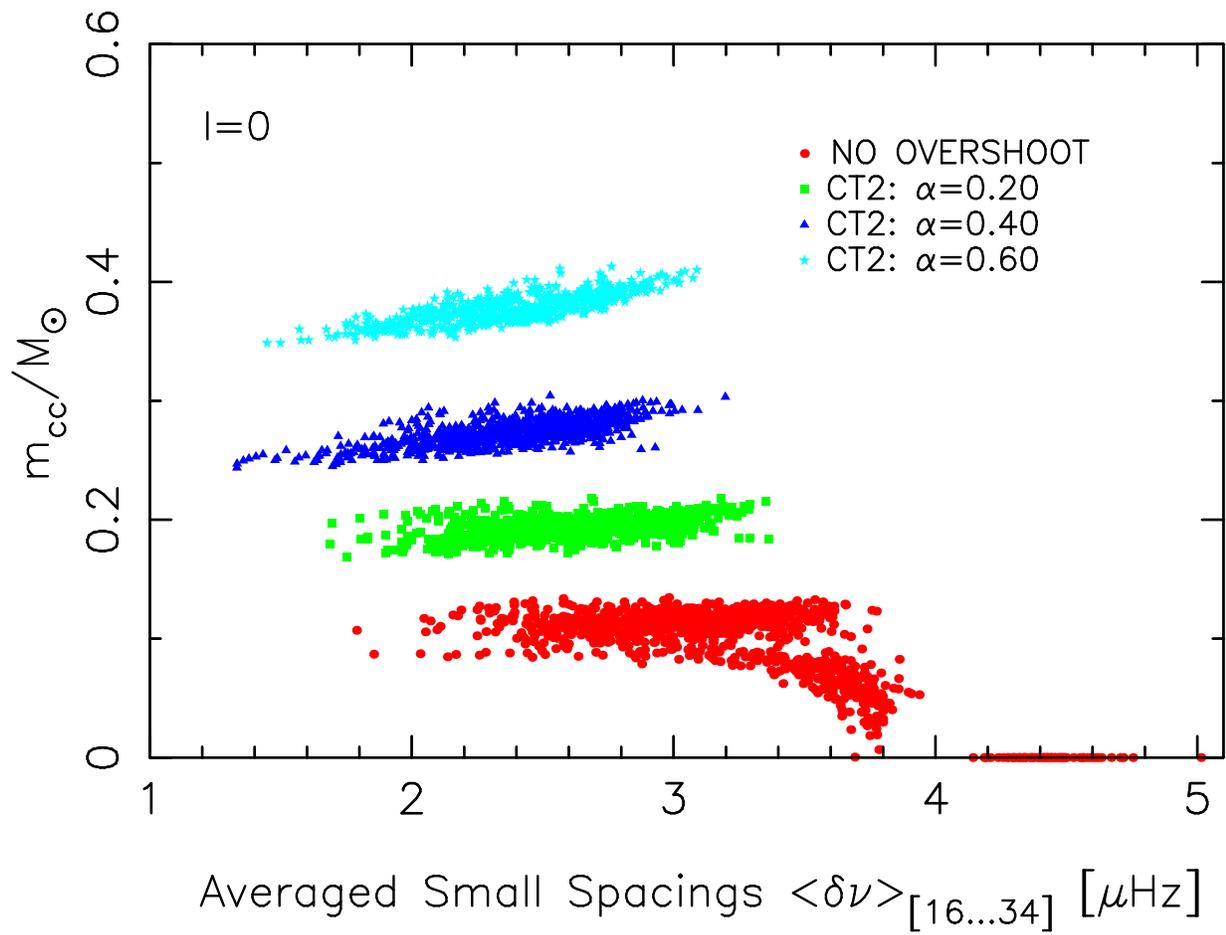}
\caption{
Averaged small p-mode spacings versus convective core size for
the $\lz$ mode and models with CT2.
\label{fig:cz_small_l0}}
\end{figure}

\begin{figure}
\includegraphics[height=\textwidth,angle=-90]{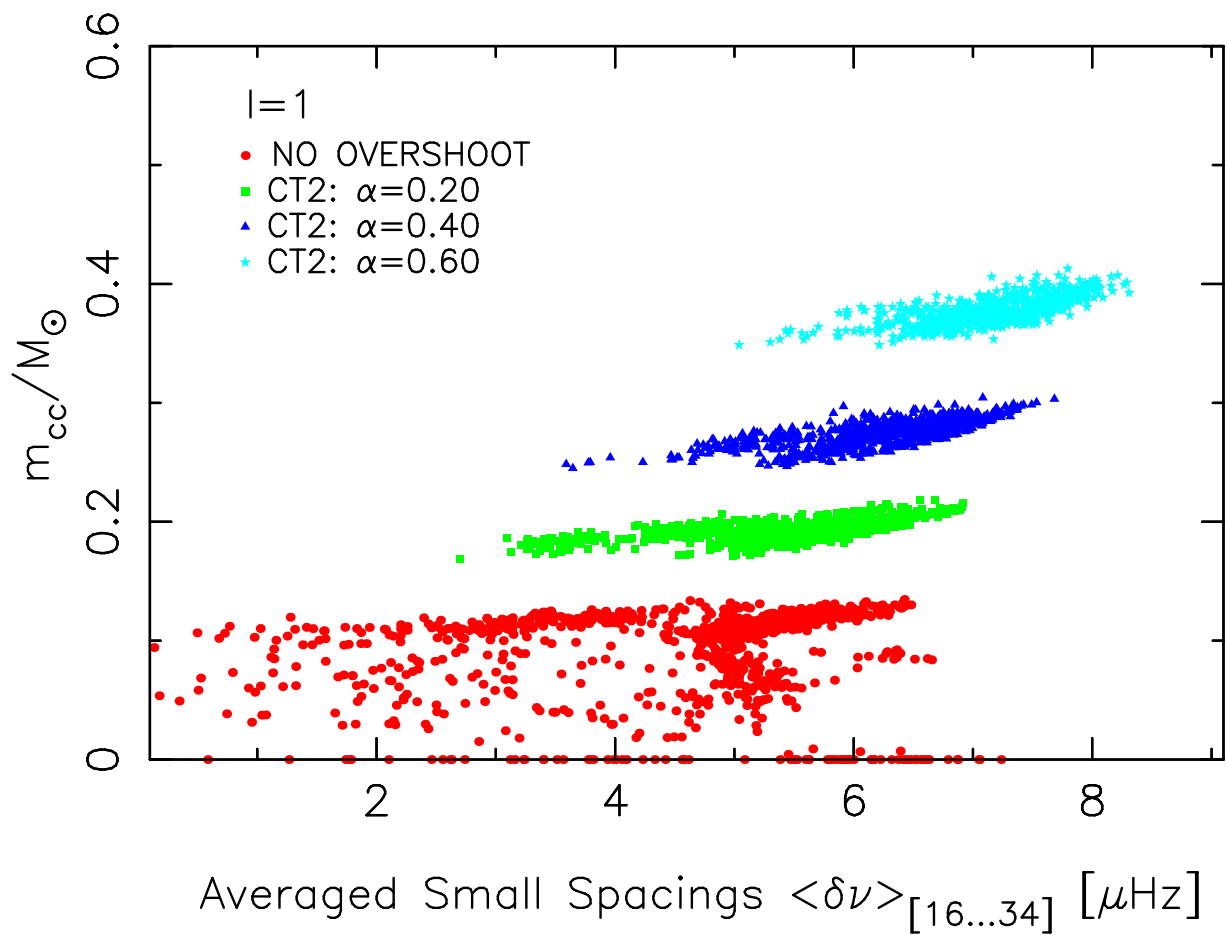}
\caption{
In comparison to the $\lz$, the $\lo$ averaged small p-mode
spacings show a better correlation with convective core size.
\label{fig:cz_small_l1}}
\end{figure}

Therefore, we want to know whether the small spacings could be used
to discriminate between different convective core sizes.
We show the small spacings versus the core size, averaged over radial
orders $n=[16 \ldots 34]$. If we look at the $\lz$, $CT2$ models
in \Fig\ref{fig:cz_small_l0}\notetoeditor{\Fig\ref{fig:cz_small_l0}
and \Fig\ref{fig:cz_small_l1} should appear side-by-side.}
we see that for frequency spacings
$>3.4\,\mu\mbox{Hz}$ only models without overshoot can be found.
In addition, almost all subgiant models which can be identified as the ones
exhibiting zero convective core sizes occupy exclusively the region
$>4\,\mu\mbox{Hz}$. This result is robust, because it does not
depend on the overshoot prescriptions since only tiny amounts of
overshoot inhibit the subgiant models completely. Thus, the averaged
small spacings for $\lz$ might provide the necessary measurement
needed to discriminate between the two evolutionary stages.

In comparison to the $\lz$ modes the $\lo$ frequency spacings are more
influenced by the different amounts of overshoot. Here, overshoot
leads to larger frequency spacings (\Fig\ref{fig:cz_small_l1}). Due to
the spreads in parameter space, inferring the overshoot parameter
or convective core size is difficult: If, for example, a
frequency spacing of $5\,\mu\mbox{Hz}$ is measured, $\alphOV$
could be anywhere between zero and $0.6$. The situation is
more favorable for a measured frequency spacing of $8\,\mu\mbox{Hz}$
where one could infer an $\alphOV\sim0.4\pm0.2$.
As will be shown in the next section, the g-modes are better
suited to yield a core size measurement.

\subsection{g-modes}
\label{sec:gmodes}
Although g-modes have never been detected conclusively in the Sun
\markcite{2000ApJ...538..401A,2003ApJ...588.1199W}({Appourchaux} {et~al.} 2000; {Wachter} {et~al.} 2003), it may be
possible to detect them in \ProcAk. The
convection zone in the sun's atmosphere constitutes a long attenuating
tunnel for g-modes \markcite{2002tsai.book.....S}({Stix} 2002, p.234) in strong
contrast to Procyon's very thin convective envelope of
$\sim 10^{-5}\Mstar$. On this basis we hope that for
existing and planned asteroseismic observations, some
effort is made to detect g-modes in \ProcA.

\subsubsection{Period Spacings}
\begin{figure}
\includegraphics[height=\textwidth,angle=-90]{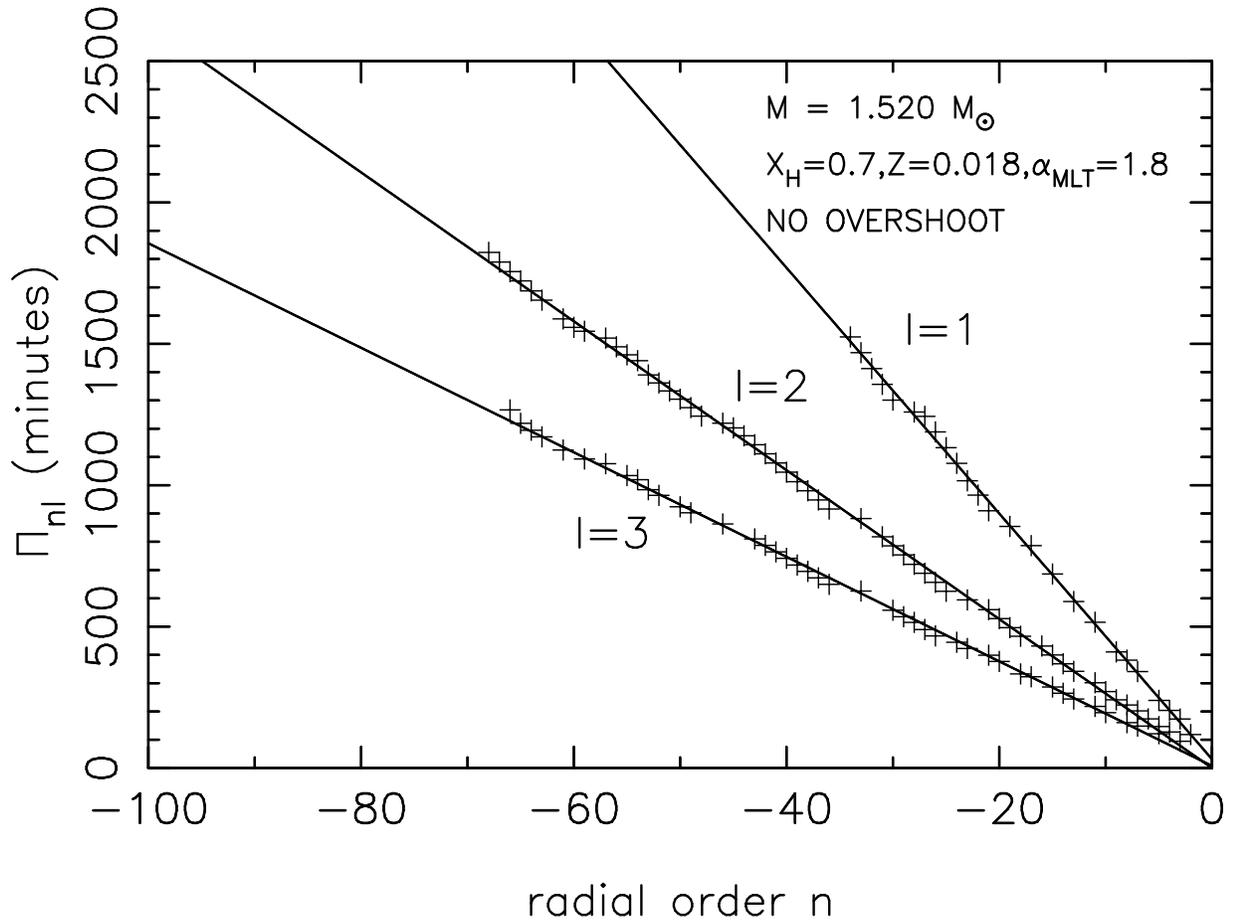}
\caption{Calculated g-mode periods as a function
of radial order $n$ from pulsation analysis for first
three $l$'s. For given $l$, the steepness of the best fit to the periods
gives the \emph{period spacings}. Most properties of the
period spacings can be explained by asymptotic theory.\label{fig:gmode1}}
\end{figure}

For all our models that fall within the error box we calculate the
$\lo,2,3$ g-modes from radial order $n=-2$ to about $-30 (\lo)$, 
$-70 (\ltw)$, $-80 (\ltr)$ respectively with the
non-radial pulsation code JIG. In the following, we use the full
set of combinations,
i.e., full input parameter space and $2\sigma$ observational error box.
As described in \Sec\ref{sec:consistOBS} we perform a pulsational
analysis on five selected models within the error box, if possible.
For one selected model, we show their periods as a function of radial
order $n$ (\Figs\ref{fig:gmode1}). 
There are a few gaps in the range of g-modes computed, an artifact of the
automated search technique used by the pulsation program. This is of no
consequence here since we are here only interested in
computing the slope of the curves.
We use a least-square algorithm to fit linear functions to the calculated
gravity modes. The steepness of these lines gives the period spacings
which can be explained by asymptotic theory
(\Sec\ref{sec:asymtheory}).
Since asymptotic theory predicts a unique period spacing any scatter
of calculated frequencies from a straight line indicates deviations from
this theory. There is some scatter at very low $n$ where asymptotic
theory does not hold and also where mode bumping is likely to occur.
In addition, as can be seen in \Fig\ref{fig:gmode1} but is
present in all calculated g-mode spectra, the scatter
is larger for lower $l$ degree modes especially at high radial order.
This scatter may well contain information about stellar structure
beyond the information contained in the period spacing but we confine
ourselves to the latter because the scatter is a quantity much more
difficult to measure.

\subsubsection{Asymptotic Theory}
\label{sec:asymtheory}

According to the asymptotic theory \markcite{1980ApJS...43..469T}({Tassoul} 1980),
the period $\Pi_{nl}$ of low
degree $l$ and high radial order $n$ ($n \gg l \simeq 1$) is, to first
order, given by \markcite{1989nos..book.....U}({Unno} {et~al.} 1989, p.83)
\begin{equation}
\Pi_{nl} \simeq \frac{(n+l/2-1/4)}{\sqrt{l(l+1)}}\, \Pi_0
\end{equation}
with
\begin{equation}
\Pi_0 = 2 \pi^2 \left[\int_{\ln(\rcz)}^{\ln(\rb)} N(r)\, d\ln r \right]^{-1}
\end{equation}
where $\rcz$ is the radius of the core convection zone,
$\rb$ the radius at the bottom of the convective envelope and
$N$ is the \BruntV frequency. This formula
establishes a regular period spacing $\Pi_{l}$ which is given by
\begin{equation}
\label{eq:Pil}
\Pi_{l} = \frac{2 \pi^2}{\sqrt{l(l+1)}} \left[\int_{\ln(\rcz)}^{\ln(\rb)} N(r)\, d\ln r \right]^{-1}
\end{equation}
As can be seen by this formula, the higher degree $l$ modes do not
hold additional information about the star's structure.
The most important contributions to the integral in \Eqn(\ref{eq:Pil})
come from the fully ionized inner regions of the star, hence the
square of the \BruntV frequency, $N^2$, is well
approximated by making use of the gradient of the mean molecular weight,
$\mu$, i.e.,
\begin{equation}
N^2 = \frac{g^2 \rho}{P} \left[ \frac{4 - 3 \beta}{\beta}
(\nablad - \nabla) + \varphi \nablmu \right]
\label{eq:bruntv}
\end{equation}
where we denote local gravitational acceleration $g=(G m)/r^2$,
density $\rho$, pressure $P$, ratio of gas to total
pressure $\beta = \Pgas/P$, adiabatic gradient
$\nablad$ and actual stellar temperature gradient $\nabla$. 
The \BruntV frequency gives the frequency by which a bubble of
gas oscillates vertically around its equilibrium position under
gravity. The frequency is therefore proportional to its restoring
buoyancy force which is given by the spatial entropy gradient, or
alternatively, by $\nablad - \nabla$. Another contribution
to the \BruntV frequency is given by the second term
in brackets of \Eqn(\ref{eq:bruntv}): The molecular weight gradient
which can increase or decrease the frequency depending on the
sign of $\nablmu$.

\subsubsection{g-mode Characteristics in \ProcA}

\begin{figure}
\includegraphics[width=\textwidth,angle=0]{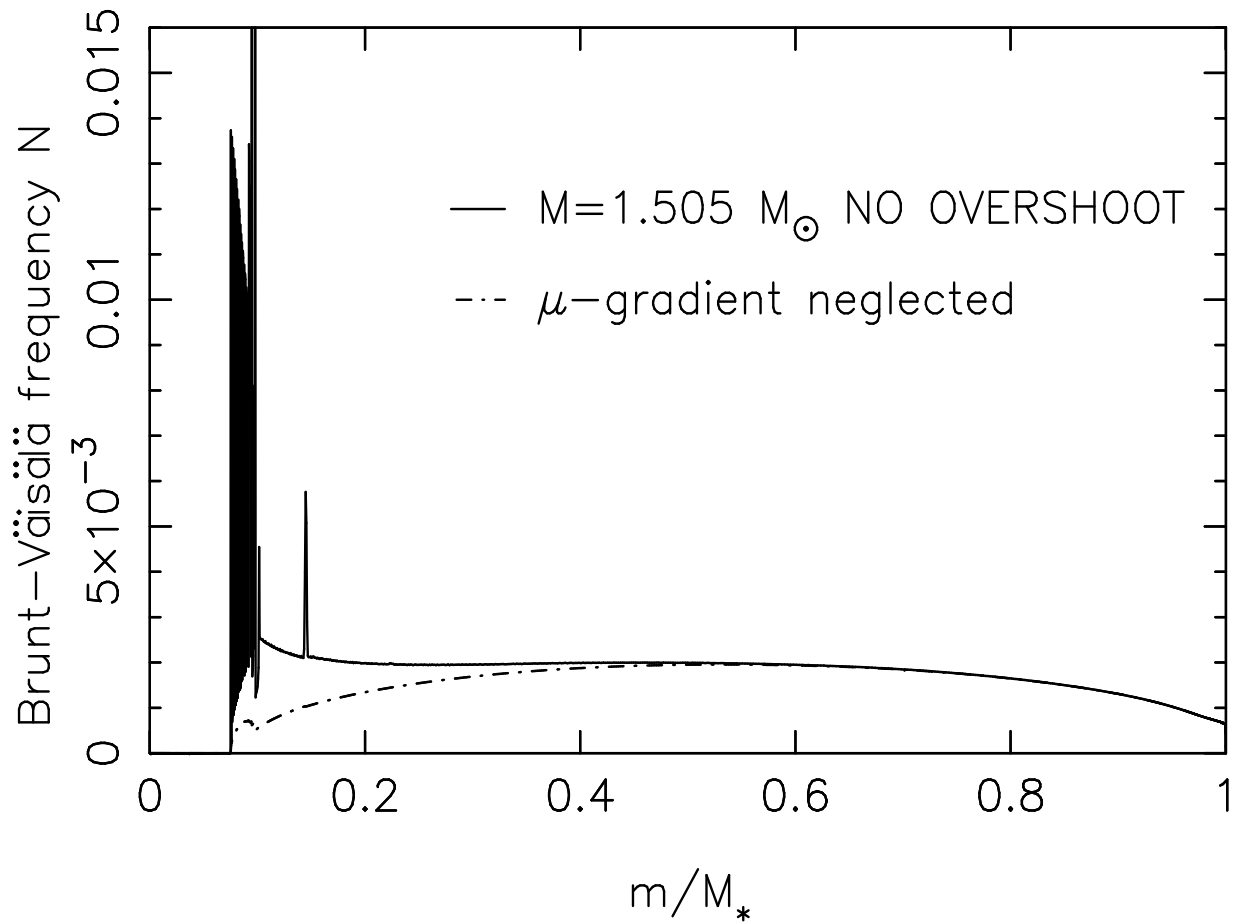}
\caption{\BruntV frequency N as a function of
stellar mass for a \ProcA model with $M=1.505\,\Msun$ without
overshoot (solid curve). To show the relative importance of the
molecular weight gradient, N is also plotted without the
contribution from $\nablmu$ (dash-dotted curve).\label{fig:bruntvex}}
\end{figure}

\begin{figure}
\includegraphics[width=\textwidth,angle=0]{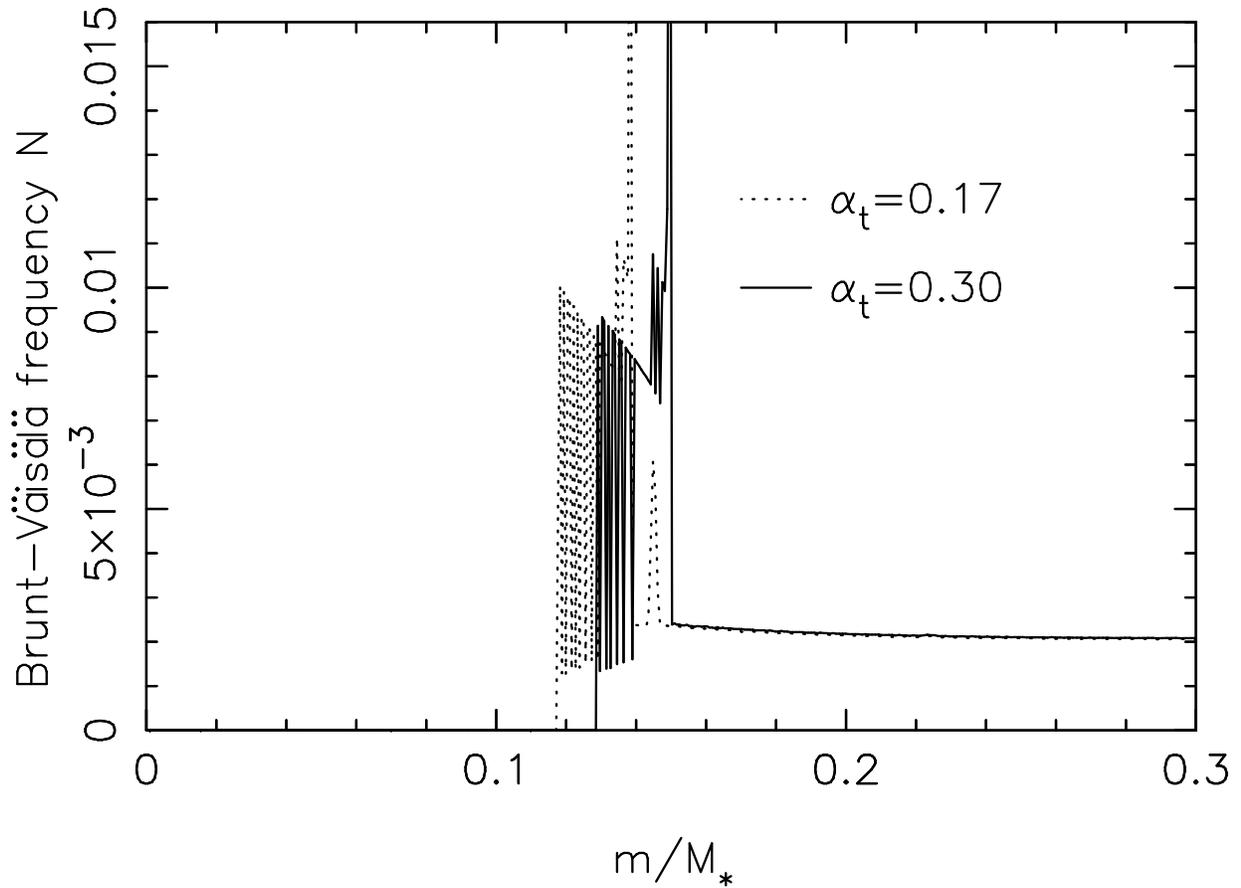}
\caption{Comparison of \BruntV frequency~N of
a model with $M=1.497 M_{\sun}$ and KCT overshoot for
two different overshoot parameters $\alpht=0.30$ (solid curve)
and $\alpht=0.17$ (dotted curve). Note that the inner
cutoff radius is larger in the case of more overshoot.\label{fig:bruntvgl}}
\end{figure}

In the case of the late main-sequence
star \ProcA the \BruntV frequency is largest
in parts of varying molecular weight gradient
which is left over from the receding core convection zone of main
hydrogen burning (\Fig\ref{fig:bruntvex}
\notetoeditor{\Fig\ref{fig:bruntvex} and \Fig\ref{fig:bruntvgl}
may appear side-by-side.}). The molecular
weight gradient for mass shells greater than about $0.4\Mstar$
remain untouched and the buoyancy starts to dominate
\Eqn(\ref{eq:bruntv}). Although buoyancy constantly increases toward
the outer stellar layers the drop-off of density reduces its
contributions to the frequency exhibiting a maximum around $0.6\Mstar$.
These general characteristics are true for all our models irrespective
of any overshoot, because this basic structure can be found in
all late main-sequence stars that have convective cores.

\subsubsection{What g-modes reveal about overshoot}

With asymptotic theory (\Sec\ref{sec:asymtheory}),
we can gain further insight on how overshoot influences
the g-mode period spacings.
In order to show the differences in the 
\BruntV frequency inscribed by different amounts of
overshoot we show \Fig\ref{fig:bruntvgl} where we compare
KCT overshoot for two different overshoot parameters
$\alpht = 0.17$ and $\alpht =0.30$. We recall, that since
overshoot increases the size of the convective region
it taps a larger hydrogen fuel reservoir.
Models which fall into the same error box and exhibit larger overshoot
have not used up hydrogen as much and therefore the
molecular weight gradient is shallower albeit increasing the
period spacing $\Pi_{0}$. Not so obvious at first glance but even
more important is the inner cutoff radius $\rcz$ which is larger
for models with overshoot, reduces the \BruntV frequency
and therefore also increases the period spacing $\Pi_{0}$.

These effects on the period spacings imply, that for
given and fixed overshoot prescription and
input parameters (mass, metallicity, etc.) a one-to-one relation
between period spacings and convective core size could be made.
On the other hand, the situation here is more complicated
since we must take a range of uncertainties into account.

\begin{figure}
\includegraphics[height=\textwidth,angle=-90]{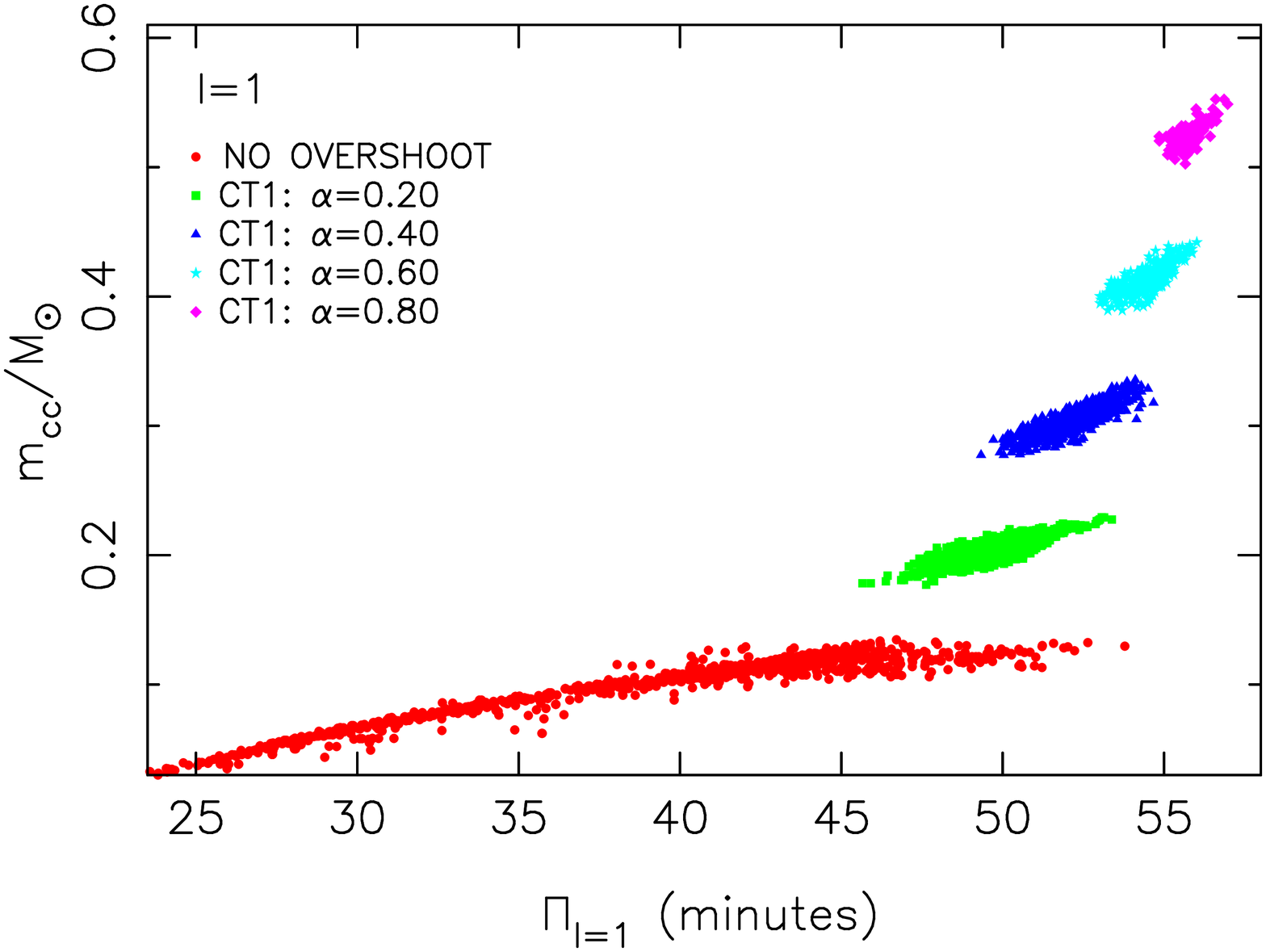}
\caption{Correlation between period spacings of g-modes
and convective core size for lowest order $\lo$ mode.\label{fig:mczpivgl1}}
\end{figure}

\begin{figure}
\includegraphics[height=\textwidth,angle=-90]{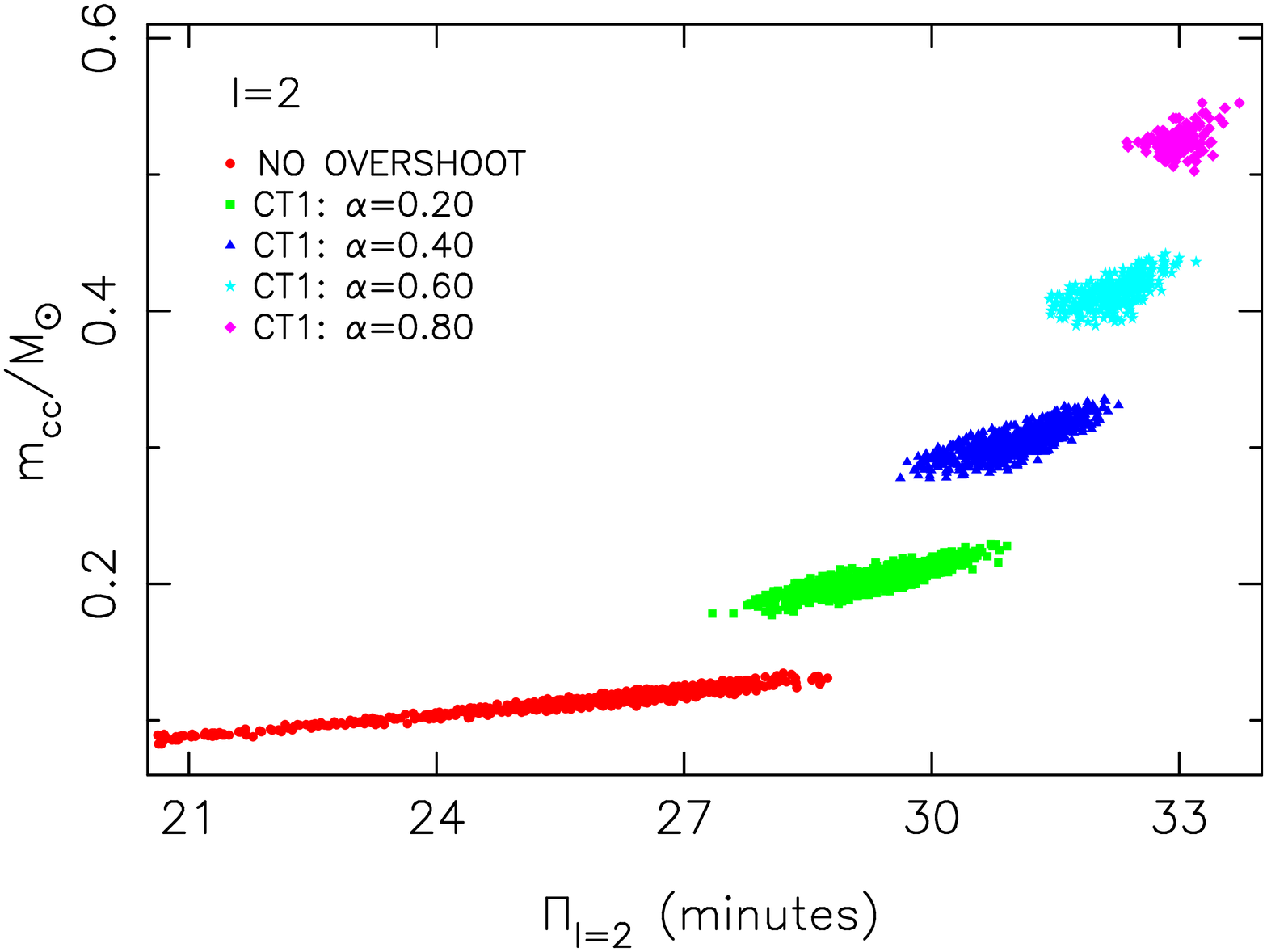}
\caption{Correlation between period spacings of g-modes
and convective core size for $\ltw$ mode.\label{fig:mczpivgl2}}
\end{figure}

\begin{figure}
\includegraphics[height=\textwidth,angle=-90]{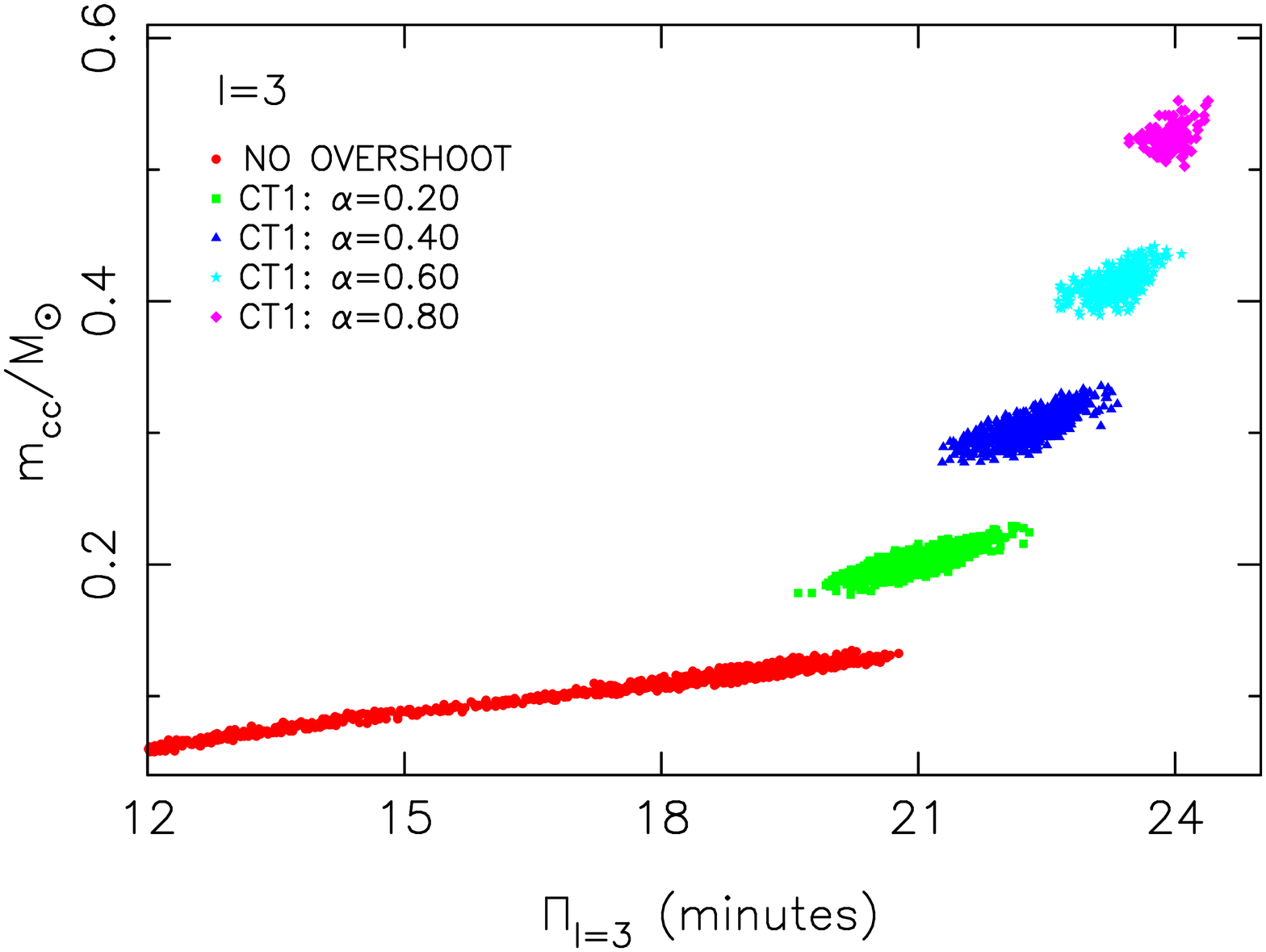}
\caption{Correlation between period spacings of g-modes
and convective core size for $\ltr$ mode.\label{fig:mczpivgl3}}
\end{figure}

\begin{figure}
\includegraphics[height=\textwidth,angle=-90]{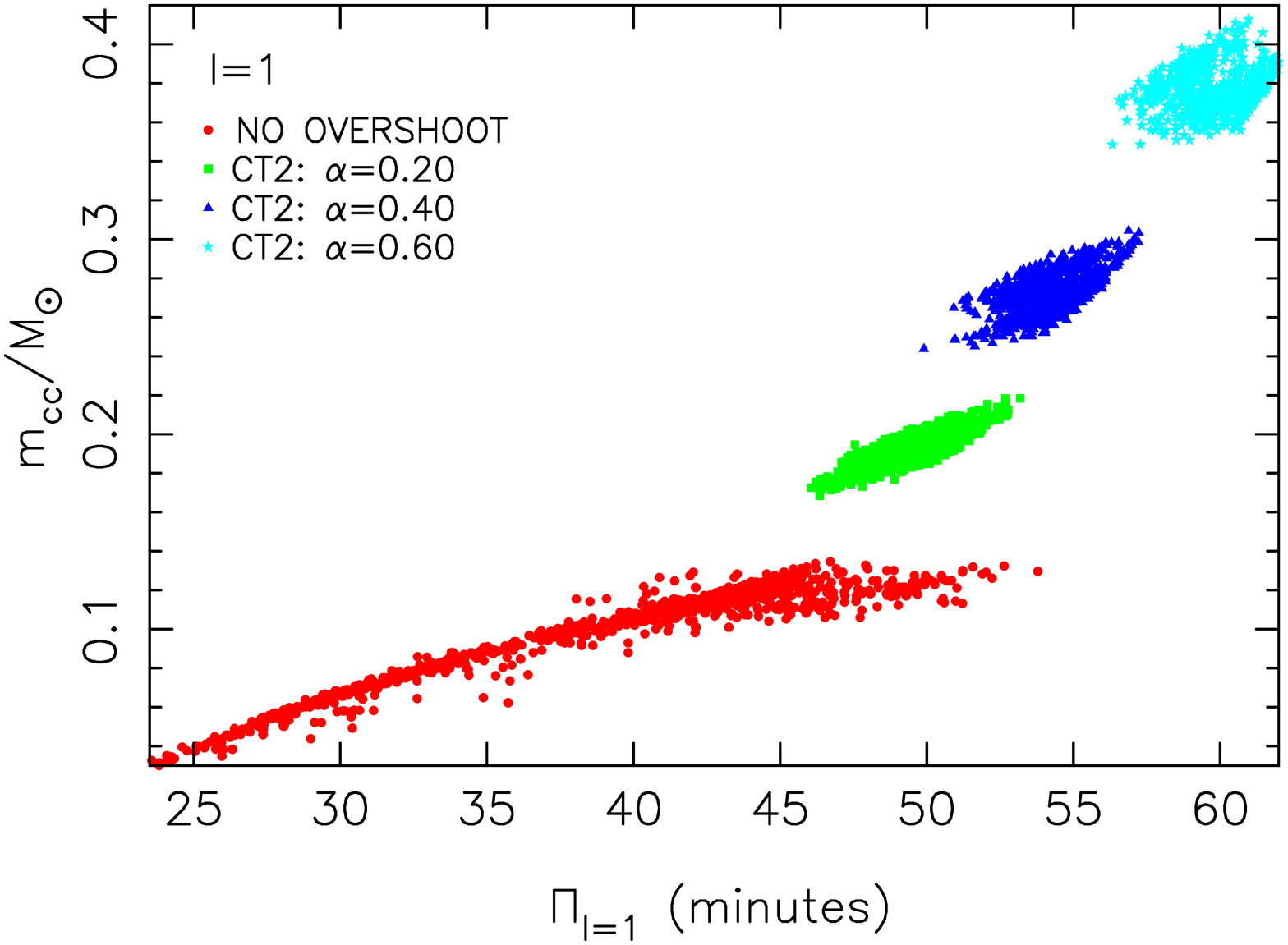}
\caption{
Same as \Fig\ref{fig:mczpivgl1} but for CT2 convection
models.\label{fig:mczpivgl4}}
\end{figure}

\begin{figure}
\includegraphics[height=\textwidth,angle=-90]{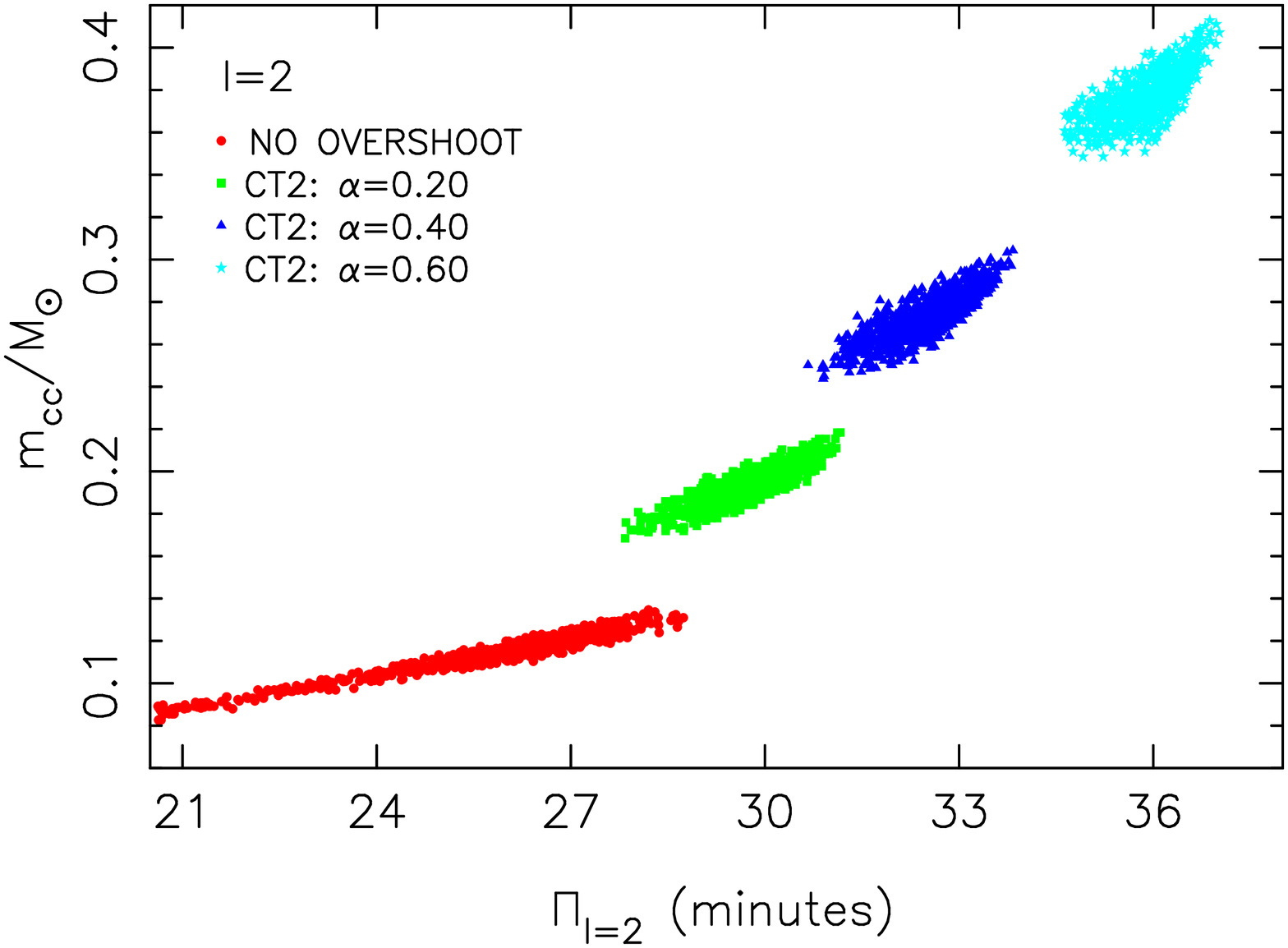}
\caption{
Same as \Fig\ref{fig:mczpivgl2} but for CT2 convection
models.\label{fig:mczpivgl5}}
\end{figure}

\begin{figure}
\includegraphics[height=\textwidth,angle=-90]{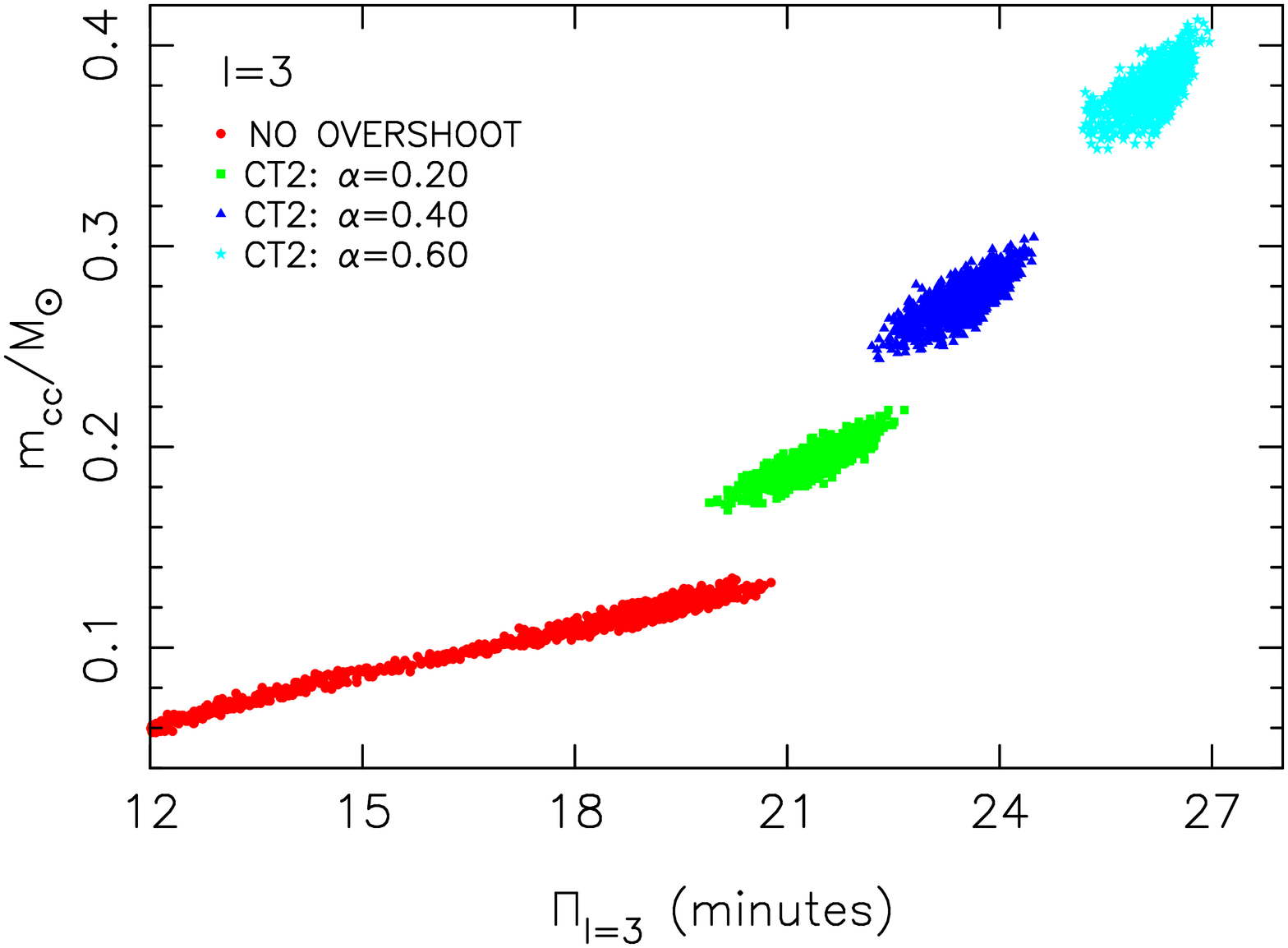}
\caption{
Same as \Fig\ref{fig:mczpivgl3} but for CT2 convection
models.\label{fig:mczpivgl6}}
\end{figure}

\begin{figure}
\includegraphics[height=\textwidth,angle=-90]{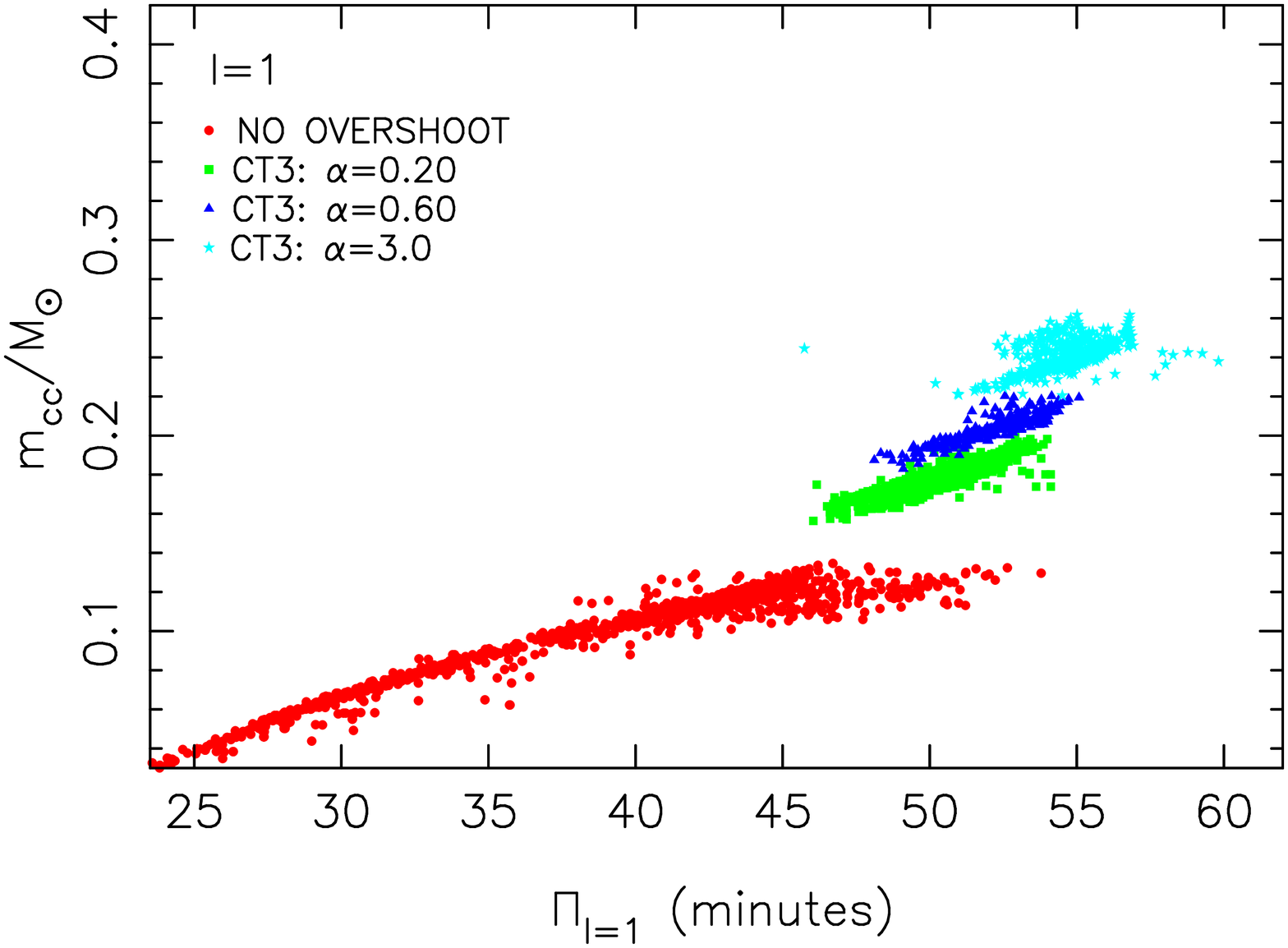}
\caption{
Same as \Fig\ref{fig:mczpivgl1} but for CT3 convection
models.\label{fig:mczpivgl7}}
\end{figure}

\begin{figure}
\includegraphics[height=\textwidth,angle=-90]{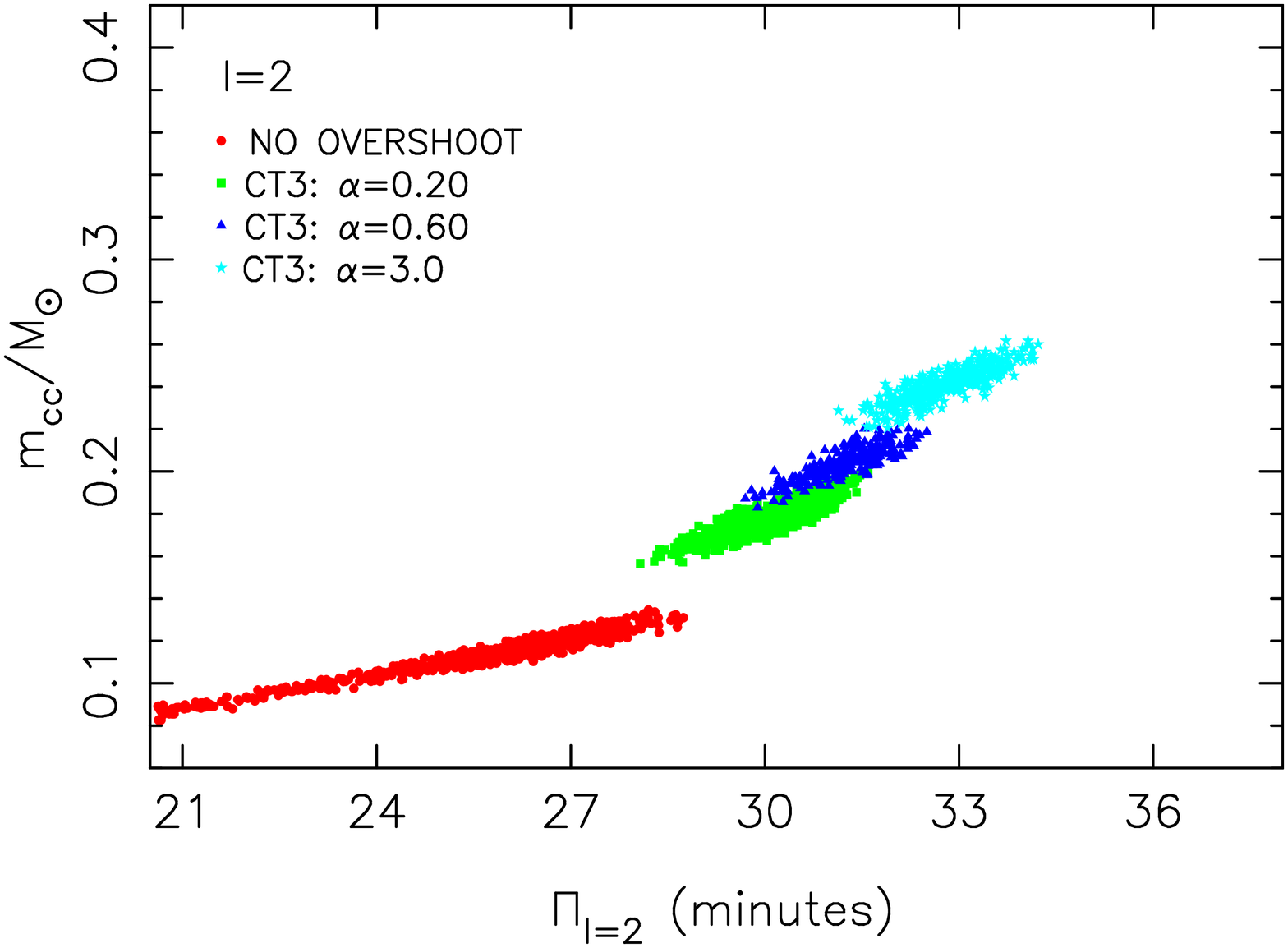}
\caption{
Same as \Fig\ref{fig:mczpivgl2} but for CT3 convection
models.\label{fig:mczpivgl8}}
\end{figure}

\begin{figure}
\includegraphics[height=\textwidth,angle=-90]{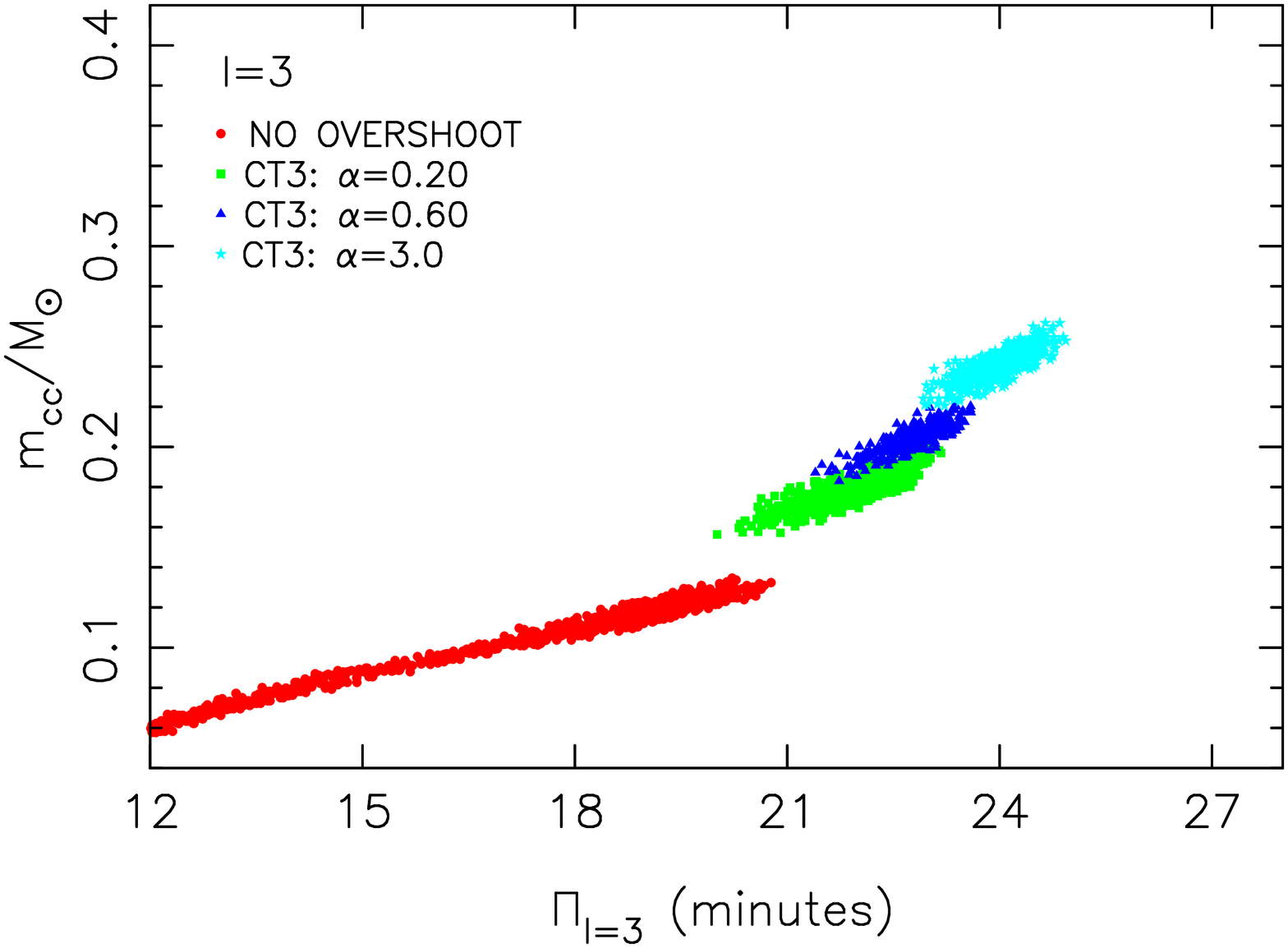}
\caption{
Same as \Fig\ref{fig:mczpivgl3} but for CT3 convection
models.\label{fig:mczpivgl9}}
\end{figure}

Here we present the relation between period spacings and
convective core size in
\Figs\ref{fig:mczpivgl1}-\ref{fig:mczpivgl9}
\notetoeditor{\Fig\ref{fig:mczpivgl1}-\Fig\ref{fig:mczpivgl3}
must be grouped together. They may appear together with
Fig\ref{fig:mczpivgl1}-\Fig\ref{fig:mczpivgl9}.}
\notetoeditor{\Fig\ref{fig:mczpivgl4}-\Fig\ref{fig:mczpivgl6}
must be grouped together. They may appear together with
Fig\ref{fig:mczpivgl1}-\Fig\ref{fig:mczpivgl9}.}
\notetoeditor{\Fig\ref{fig:mczpivgl7}-\Fig\ref{fig:mczpivgl9}
must be grouped together. They may appear together with
Fig\ref{fig:mczpivgl1}-\Fig\ref{fig:mczpivgl9}.}
for all models of the full input parameter
set that fall in the $2\sigma$ observational error box. We
can see a strong albeit not perfect {\it correlation of the
period spacing with convective core size}.  This can be used
to measure the overshoot parameter within a given overshoot
prescription provided it is possible to derive the period
spacing from observational data. However, within the current
uncertainties, this measurement is not extremely
accurate. For adopted overshoot prescription and given value
of the period spacings the derived overshoot parameter
can be determined with an accuracy of $\pm 0.15\,\Hp$
(CT1), $\pm 0.10\,\Hp$ (CT2) and $\pm 0.05\,\Hp$ (CT3). Note
that although the overshoot parameter $\alpht$ can only be
constrained within $\pm 1.4$ the overshoot extend measured
in pressure scale heights is most favorably constrained with
CT3.  We will see in the next \Sec\ref{sec:sensitivity} what
uncertainties need to be improved upon in order to be able
to get a more constrained overshoot estimate.

Still, we can deduct valuable information about the
convective core size, especially in certain period regions.
Whereas models without overshoot occupy the regions
with period spacings of $25-45 (\lo)$, $21-27 (\ltw)$ and $12-19 (\ltr)$
minutes the models with overshoot have generally longer period spacings.
Thus we find, that irrespective of the different overshoot
prescriptions, there exists a period regime for models without overshoot
that is cleanly separated from the overshoot cases.

For period spacings greater than $45 (\lo)$, $27 (\ltw)$ and $19 (\ltr)$
we get two different types of overlaps. First, for given adopted overshoot
prescription and overshoot parameter, there is a spread due
to the input model and observational uncertainties. As seen in
\Figs\ref{fig:mczpivgl1}-\ref{fig:mczpivgl9} these spreads constitute
islands which correspond to a period regime which overlaps with neighboring
ones. The less overlap the more accurate it is possible
to determine the overshoot parameter from the period spacings. Note
that the overlap is stronger for the CT1 models when compared to the CT2
ones. The second overlap stems from the fact that we do not know the
correct overshoot theory before hand. For example, if we look
at the $l=2$ plots (\Figs\ref{fig:mczpivgl2},\ref{fig:mczpivgl5})
at a period of 31 minutes with CT1 we get $\alphOM=0.35\pm0.15$,
on the other hand, with CT2 we determine $\alphOV=0.3\pm0.10$.

Both types of overlaps could be reduced. The first one could
be reduced by better observational data that is able to narrow down
the error box and the set of input parameters. The most crucial
observational parameters for this will be identified in the next
\Sec\ref{sec:sensitivity}. The second type of
overlaps that result from the different overshoot prescriptions is
even harder to overcome: when the outlined analysis presented
here for \ProcA is applied to many different, at least equally good
constrained stars, it might be possible to do so in the future.

%To complete this section we provide the period spacings calculated in
%our pulsation analysis for $\lo,2,3$ for the complete range of
%models studied (\Tab\ref{tab:spacings}). We also list the
%period spacings derived from asymptotic theory which
%deviate from the ``real'' spacings about $4\%$ for $\lo$, $3\%$ for $\ltw$
%and $1\%$ for $\ltr$. Apparantly, the spacings from asymptotic
%theory are systematically higher compared to the non-radial pulsation
%analysis. These systematics are expected and described for
%polytropic models in the original work by \markcite{1980ApJS...43..469T}({Tassoul} 1980).
%The difference stems from two approximations in asymptotic theory:
%the first neglects perturbations of the gravitational potential 
%and the second is introduced when constructing the asymptotic
%expansion itself.

\subsection{Sensitivity of Results}
\label{sec:sensitivity}
In the last section we demonstrated that, for a given convection
prescription, uncertainties on our models imposed by both
the observational error box and the possible initial input
parameter space lead to overlaps in the period spacings thereby
imposing uncertainties on future measurements of the
overshoot parameter. For upcoming observations on \ProcA it might
be enlightening which parameters need to be constrained the most
in order to be able to determine the overshoot parameter with
more accuracy.

\begin{figure}
\includegraphics[height=\textwidth,angle=-90]{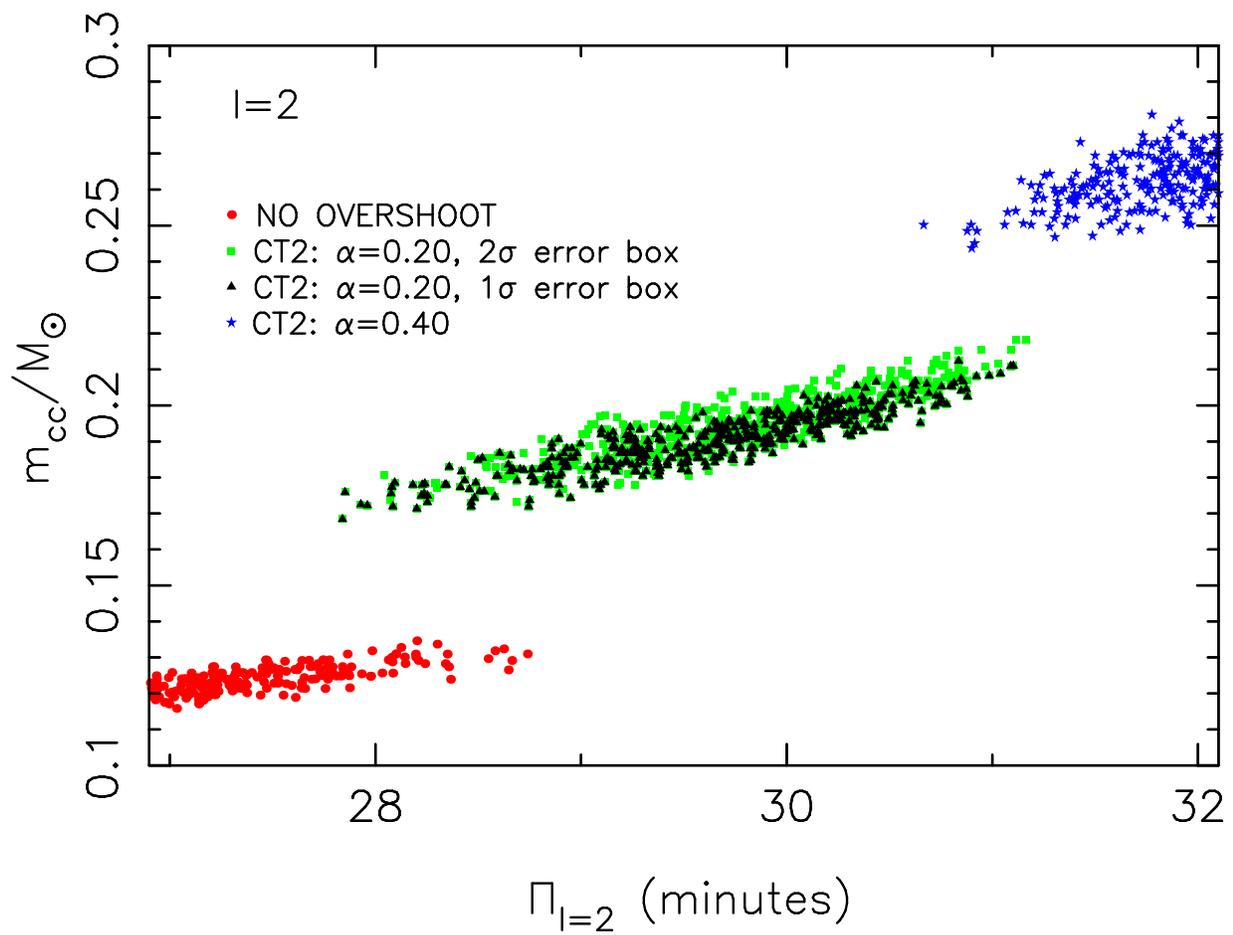}
\caption{
Model spreads for full parameter range (squares) in comparison
to models that are confined to the $1\sigma$ observational
error box (triangles).\label{fig:eboxtest}}
\end{figure}

\begin{figure}
\includegraphics[height=\textwidth,angle=-90]{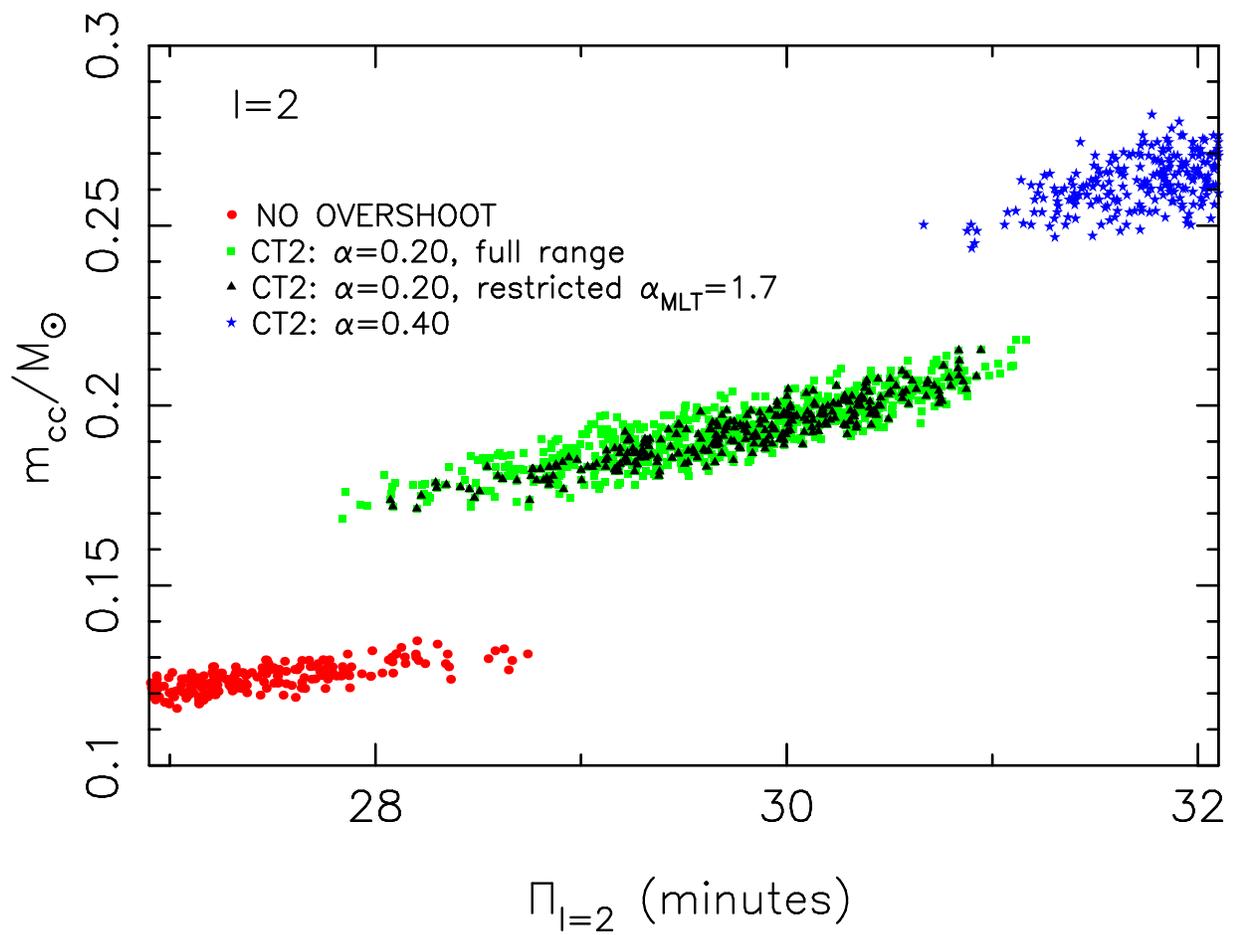}
\caption{Model spreads for full parameter range (squares)
in comparison to models with fixed mixing length
(triangles).\label{fig:ratest}}
\end{figure}

First we want to determine the effects of narrowing the
observational error box. To do so, we only allow those models to
qualify as viable models that fall into the $1\sigma$ observational
error box. As seen in
\Fig\ref{fig:eboxtest}\notetoeditor{\Fig\ref{fig:eboxtest}-\Fig\ref{fig:ratest}
should be grouped together.} the period spread
occupied by the $CT2$, $l=2$, $\alphOV=0.2$ \emph{$1\sigma$ error box}
models (triangles) is not smaller compared to the
full model set (squares). Thus, narrowing the error box to
$1\sigma$ does not help in improving the accuracy of the
overshoot parameter.

\begin{figure}
\includegraphics[height=\textwidth,angle=-90]{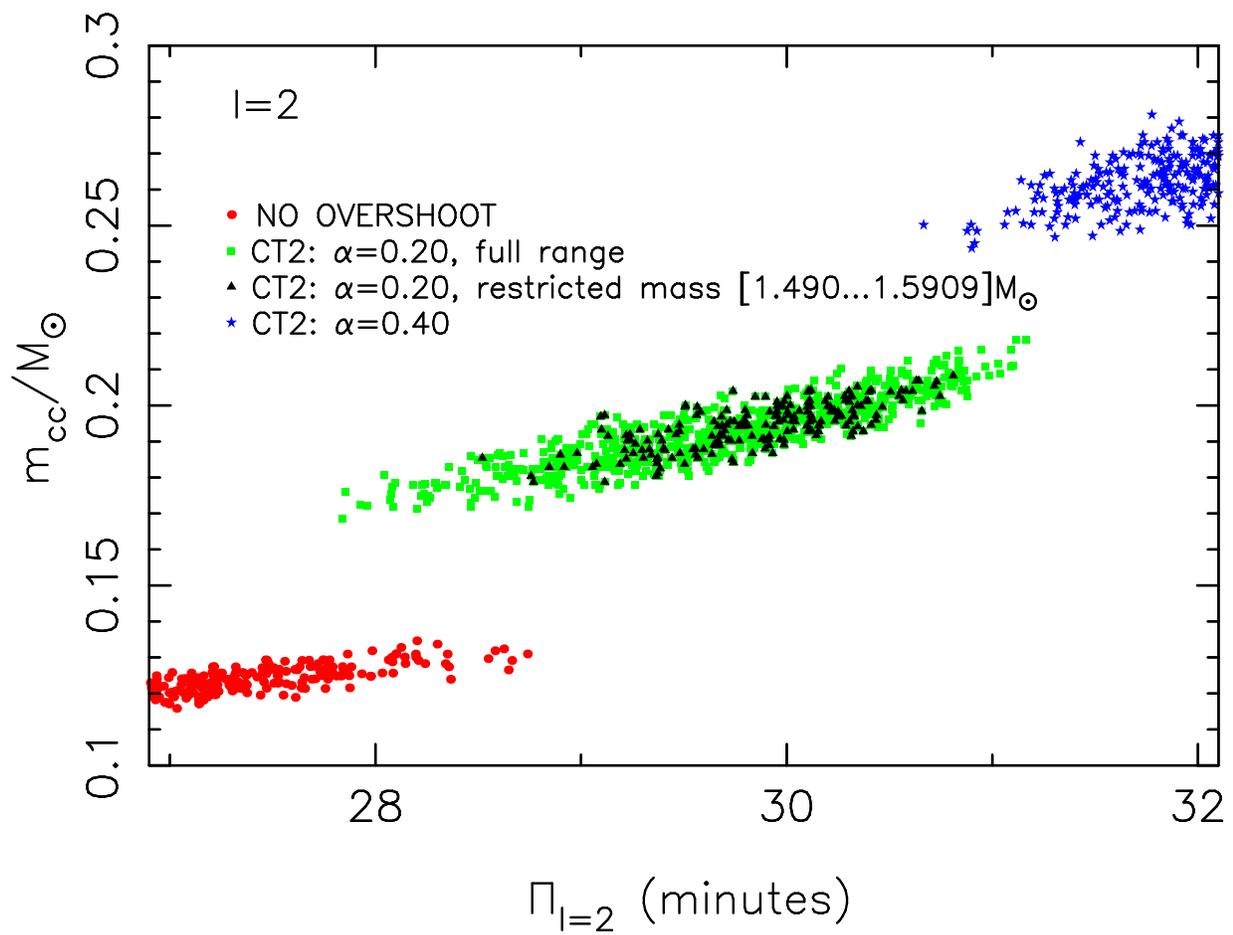}
\caption{Model spreads for full parameter range (squares) in comparison
to models with restricted mass (triangles).\label{fig:rmtest}}
\end{figure}

\begin{figure}
\includegraphics[height=\textwidth,angle=-90]{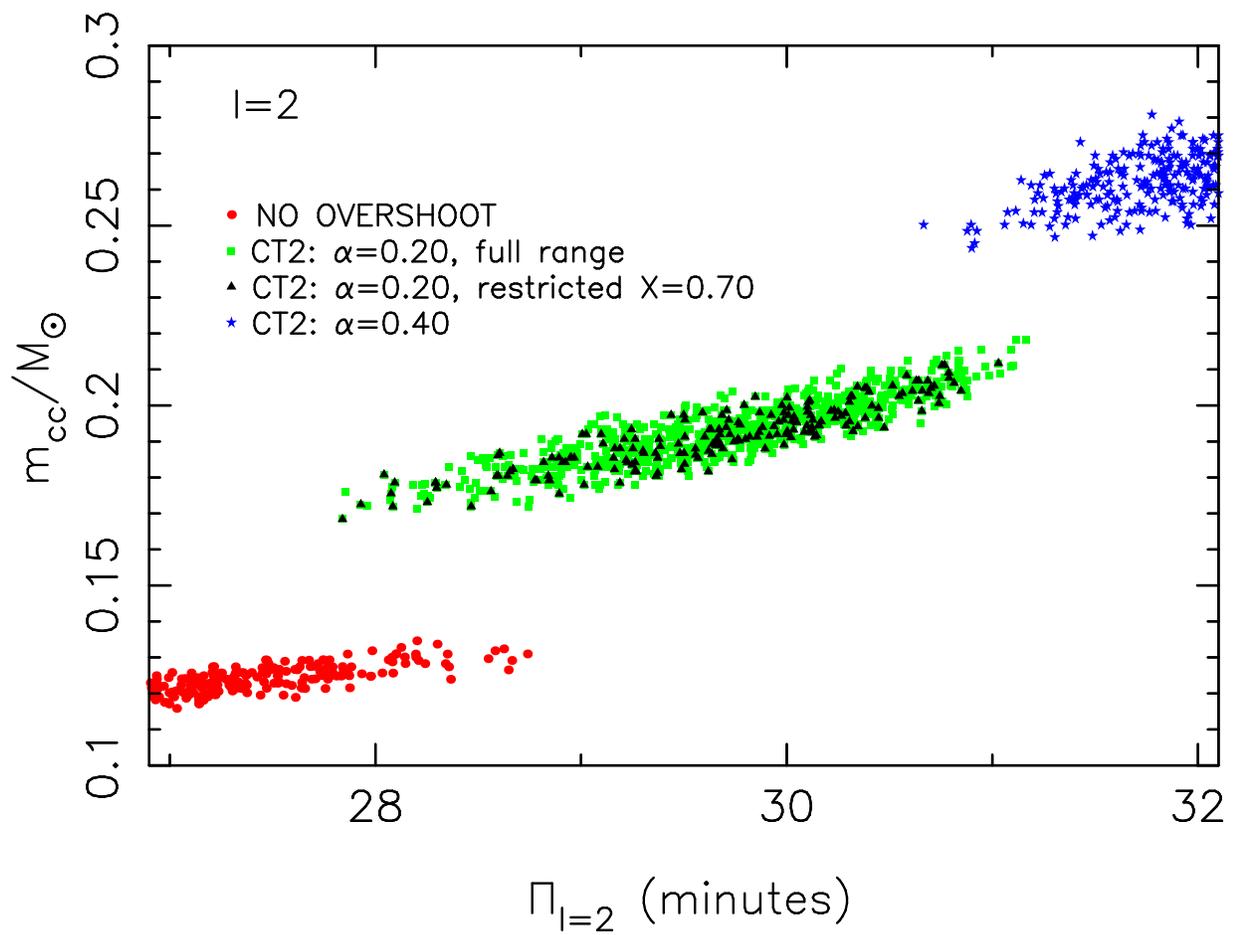}
\caption{Model spreads for full parameter range (squares)
in comparison to models with fixed hydrogen content
(triangles).\label{fig:rXtest}}
\end{figure}

\begin{figure}
\includegraphics[height=\textwidth,angle=-90]{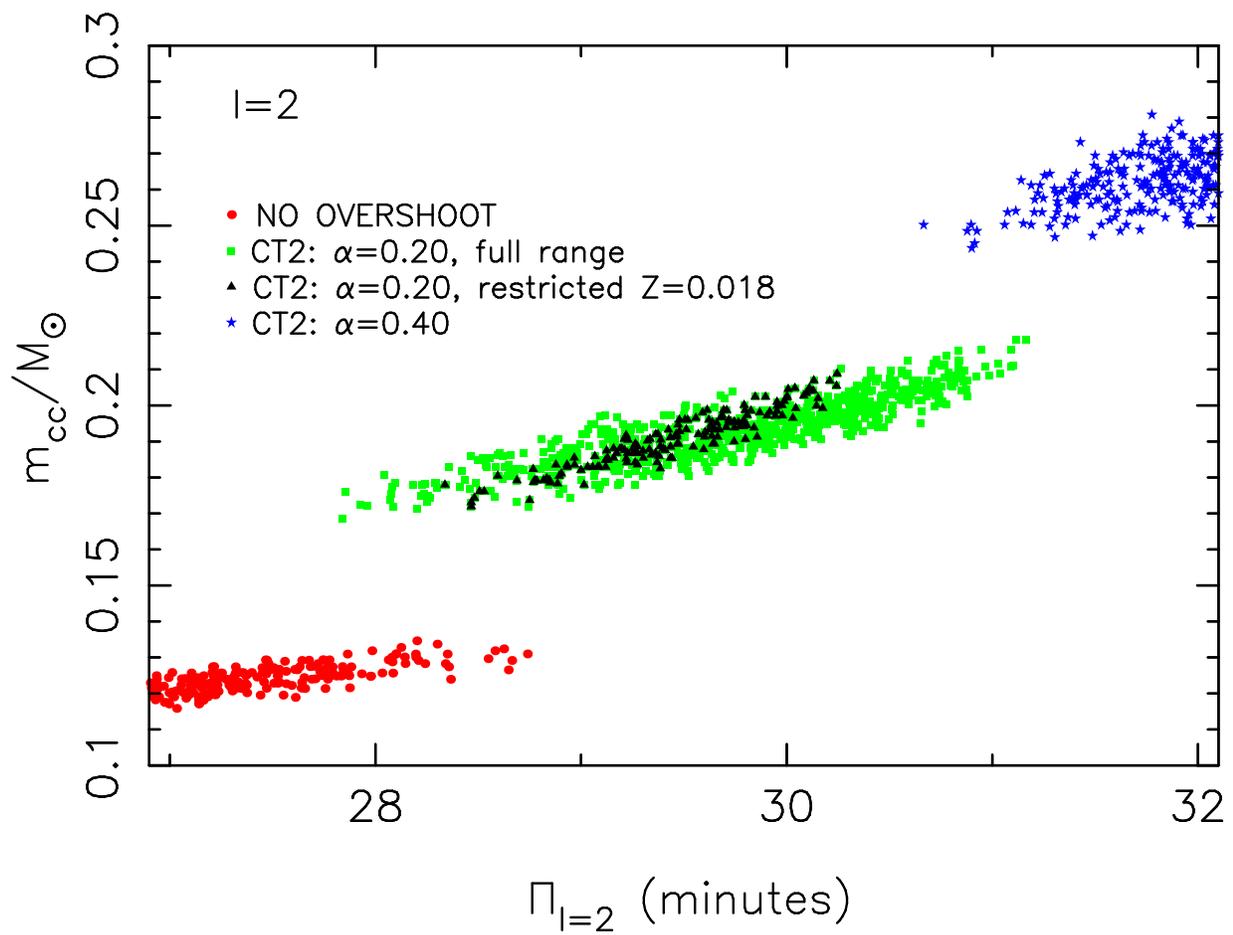}
\caption{Model spreads for full parameter range (squares)
in comparison to models with fixed metallicity
(triangles).\label{fig:rZtest}}
\end{figure}

Next we assess the possible benefits arising from
constraining one out of the five input parameters, i.e., mass
(\Fig\ref{fig:rmtest}), hydrogen fraction (\Fig\ref{fig:rXtest}), 
metallicity (\Fig\ref{fig:rZtest}) and mixing length parameter
(\Fig\ref{fig:ratest}). As can be seen in the \Figs the
most crucial input parameters turn out to be both the
mass and the metallicity. Measuring those two more precisely
can bring down the spread by a factor of three.

\section{Summary and Conclusions}
We have constructed a comprehensive set of stellar models for \ProcA
with varying model assumptions on mass, hydrogen content,
metallicity, mixing length parameter,
adopted core overshoot prescription and overshoot parameters
in an effort to single out current and promising future observables
that constrain the amount and possibly the nature of core overshoot
in stellar interiors.

In this study, effects of rotation have not been taken into account.
\ProcA is a slow rotator with a measured rotational velocity
of $v \sin i = 3.2\, \km\,\mbox{s}^{-1}$. Rotation is
therefore expected to produce multiplets for
higher azimuthal modes than the fundamental with splittings
on the order of $0.5\,\mu\mbox{Hz}$. This has to be kept in mind
when comparing the predictions made here with actual observations.

The classical treatment of overshoot in stars ---
extending the core size by some fraction of the pressure scale height ---
is merely a measure of overshoot and cannot be regarded as
an accurate description of the underlying physics. For this
basic reason we implement a physically better motivated
non-local convection theory developed by \markcite{1986A&A...160..116K}{Kuhfu{\ss}} (1986).
We neglect a number of terms present in this theory, foremost
its time-dependency and some others expected to be only
important in evolutionary phases that must be followed on a
dynamical timescale. More relevant to this study is the
choice of the five free parameters present in the theory. Reasonable
choices on four can be made in analogy with MLT and we only
vary one parameter, $\alpht$, which is the most relevant to the
amount of overshoot. However, we stress that all parameters must
be eventually calibrated, a task that might prove possible
in the future if the methods presented in this paper are applied
to many stars alike. Regardless of the actual adopted numbers,
we do find a moderate increase of the amount of core overshoot with
increasing stellar mass up to about $15\,\Msun$. In comparison
with Kuhfu{\ss}'s original analysis we find smaller amounts
of overshoot for the same adopted parameters mainly because
we limit the pressure scale height in the stellar center.
We are not certain, whether the implementation by
\markcite{1987PhDT.......162K}{Kuhfu{\ss}} (1987) omitted the molecular weight gradient
which would constitute an additional reason why he found
overall larger convective core sizes.

We also show, that models calculated with the Kuhfu{\ss} convection
theory contain convective core overshoot extents
smaller than $\sim 0.4\,\Hp$
regardless of the overshoot parameter $\alpht$ adopted. This
behavior is rooted in the non-local treatment of the theory: Since
overshooting eddies transport internal energy farther out it also
lowers the internal energy in deeper regions of the core. But
the deep regions provide the driving necessary for overshoot to occur.
Thus we have two competing effects which lead to a limit for
the extent of the core size and this limit matches
the limit imposed by Roxburgh's integral criterion.

The current observational data of \ProcA for its basic stellar
parameters of mass, luminosity, radius and effective temperature
constrain the amount of core overshoot in our models only marginally.
Models without overshoot are in accord with
the observational error box and possible input parameters. Our
strong upper limits based on traditional stellar modeling for
overshoot are $\alphOM  < 1.18$ and  $\alphOV  < 1.13$. By
restricting the input parameters of mass, hydrogen, metallicity and
mixing length to their $1\sigma$ levels we also
determine weaker upper limits, i.e., $\alphOM  < 0.65$
and  $\alphOV  < 0.75$. The convective core sizes of all
models with Kuhfu{\ss} convection are much more constrained by
the theory itself and not the observations.

\ProcA could be in either the late main-sequence or the subgiant
phase. However, only small amounts of overshoot of $\alphOM > 0.05$
exclude the subgiant models as a viable possibility. We stress
that the subgiant phase might be identified by the
$\lz$ averaged small p-mode spacings. The identification of
\ProcA as a subgiant would in turn yield an extremely precise
upper overshoot limit of $\alphOM \le 0.05$.

%pmodes
Additional information might be extracted from averaged pulsation data.
In certain frequency regions, the spread of model uncertainties
is narrow enough to use the averaged large $\lz$ spacings of the p-modes
to improve on the currently known radius of Procyon. Improving the radius
measurement by a factor of two will not lead to better estimates on the
convective core size, since the largest uncertainties reside
within the mass and metallicity parameters which we were able to
show in our sensitivity study on the g-modes. Certain frequency
regions of the averaged small spacings are exclusively
occupied by models without overshoot and in addition,
the subgiant models are separated in the $\lz$ spacings. The
p-mode separations
due to overshoot are too small to give good estimates on the
convective core size. Extremely large overshoot might be
detected, e.g., $\alphOM \sim 0.6\pm 0.2$.

%gmodes
More direct measures of overshoot are hidden in the period spacings
of g-modes. In comparison to the p-modes, the separation of
averaged period spacings due to different extents of
overshoot is more pronounced, especially for the $\lo$ modes. When
adopting $CT2$ the best possible determination of the core overshoot
extent is within $\pm0.1\,\Hp$. For $CT3$ the overshoot extent
can be determined even more accurate within $\pm0.05\,\Hp$.
Again, most models without overshoot and
in addition the subgiant models are nicely separated from the
overshoot models. Since the $CT3$ models have restricted core
sizes, this theory could be falsified for very large period spacings.

Apart from these described certain regions, considerable confusion
is added due to the different possible overshoot descriptions.
At the moment, neither
can we discriminate between the different theories nor can we
get precise measurements of the overshoot parameter based on the
averaged spacings alone. To improve on this situation we must
bring down the model spreads by getting more precise values
for the traditional stellar parameters, the most important in this
case are the mass and the metallicity. The confusion due to
different convection theories may be removed by similar
analysis for a couple of equally well selected stars.

\acknowledgments

We kindly thank R.~Ehrig from the \emph{Konrad-Zuse-Zentrum f\"ur
Informationstechnik Berlin (ZIB)} for providing and supporting us
with his code LIMEX that is used in our calculations to find the
equilibrium of KCT  more efficiently. C.W.S.\
is grateful for the hospitality he received at the
Yale Astronomy Department during his visit and
acknowledges support from the \emph{German Forschungsgemeinschaft, DFG}
(SFB 439 Galaxies in the Young Universe) and the stipend of
the \emph{Elite\-f\"order\-programm f\"ur Postdoktoranden der
Landesstiftung Baden-W\"urttemberg}. This research has been
supported in part by NASA grant NAG5-13299 (P.D.) and
an operating grant from NSERC (D.B.G.).

\appendix

\section{Derivation of Implemented Convection Equations}
\label{apx:algebra}
First, we derive an expression for the
actual temperature gradient $\nabla$ which can be given
as a function of specific turbulent kinetic energy $\wturb$. The
total flux $l/(4 \pi r^2)$
is given by the sum of convective fluxes $\jw$ and
$\jt$ and radiative flux $\Frad$,
\begin{equation}
  \frac{l}{4 \pi r^2} = \jw + \jt + \Frad \punkt
\end{equation}
With the definition for the radiative gradient $\nablrad$ and the
flux $\Frad$
\begin{eqnarray}
\nablrad &=& \frac{\Hp}{\krad T}\, \frac{l}{4 \pi r^2} \\
\Frad    &=& \frac{\krad T}{\Hp}\, \nabla \\
\krad    &=& \frac{16 \sigma T^3}{3 \kappa \rho}
\end{eqnarray}
we can eliminate the luminosity $l$ and the radiative flux $\Frad$
yielding
\begin{equation}
\frac{\Hp}{\krad T}\, \left( \jw + \jt \right) = \nablrad -\nabla \punkt
\end{equation}
With the expressions for $\jw$ (see \eq[\ref{eq:jw}]) and $\jt$
(see \eq[\ref{eq:jt}]) and the following abbreviations:
\begin{eqnarray}
a &=& \frac{\rho\, \Lambda}{\krad} \\
b &=& \alphs\, \Cp \\
c &=& b \frac{\varphi}{\delta} \nablmu \\
d &=& \frac{\alpht \Hp}{T} \left(\frac{4 \pi r^2 \rho}{\delM}\right) \\
F &=& a b \\
G &=& d/b \\
H &=& c/b \\
x_{\wturb} &=& 1 / (1 + F\, \wturb^{1/2})
\end{eqnarray}
we can solve for $\nabla$ yielding:
\begin{equation}
\nabla = \nablad + x_{\wturb}\, (\nablrad-\nablad) + (1-x_{\wturb})
         \, \left( G \gradn{\wturb} + H \right) \punkt
\end{equation}
This expression can now be substituted into the original
equation for the turbulent kinetic energy
(see \eq[\ref{eq:wturbeqnsimp}]) so we can arrive at an equation
in which the coefficients depend only on known stellar parameters.
With the additional abbreviations:
\begin{eqnarray}
I &=& \alphs \Lambda \nablad \Cp T / \Hp^2 \\
A &=& I \, \left[(\nablrad - \nablad) - \frac{\varphi}{\delta} \nablmu \right] \\
B &=& c_D / \Lambda \\
C &=& (2/3)\, G\, I \\
D &=& 1/\delM \\
E &=& (2/3)\, {(4 \pi r^2 \rho)^2 \alpht \Lambda} / \delM
\end{eqnarray}
we can give the equation for the specific turbulent kinetic energy in its final
form that we implemented in our code:
\begin{eqnarray}
  \lderivt{\wturb} &=&   A\, x_{\wturb}\,{\wturb}^{1/2} - B\,{\wturb}^{3/2} \nonumber\\
  && + C\, (1-x_{\wturb}) \gradn{\wturb^{3/2}} + D\, \gradn{}\!\left(E\, \gradn{{\wturb}^{3/2}}\right) \punkt \nonumber \\
  \label{eq:APXwturbeqmath}
\end{eqnarray}
The coefficients $A$--$E$ are functions of the stellar variables
$\nablad$, $\nablrad$, $\nablmu$, $m$, $r$, $\rho$, $T$, $\krad$, $\Hp$,
$\Cp$, $\delta$ and $\varphi$.

\bibliography{}

\clearpage

\clearpage

\clearpage

\clearpage

\clearpage

\clearpage

\clearpage

\clearpage

\clearpage

\clearpage

\clearpage

\clearpage

\clearpage

\clearpage

\clearpage

\clearpage

\clearpage

\clearpage

\clearpage

\clearpage

\clearpage

\clearpage

\clearpage

\clearpage

\clearpage

\clearpage

\end{document}